\documentclass[sn-mathphys,Numbered]{sn-jnl}

\usepackage{graphicx}
\usepackage{multirow}
\usepackage{amsmath,amssymb,amsfonts}
\usepackage{amsthm}
\usepackage{mathrsfs}
\usepackage{xcolor}
\usepackage{textcomp}
\usepackage{manyfoot}
\usepackage{booktabs}
\usepackage{algorithm}
\usepackage{algorithmicx}
\usepackage{algpseudocode}
\usepackage{listings}
\usepackage{subcaption}
\usepackage{upgreek}
\usepackage{adjustbox}
\usepackage{multirow}
\usepackage{graphicx,bbold}

\newcommand{\vc}[1]{\mathbf{#1}}
\newcommand{\mbf}[1]{\mathbf{#1}}
\newcommand{\trm}[1]{\textrm{#1}}

\newcommand{\tsf}[1]{\textsf{#1}}
\newcommand{\tabref}[1]{Tab. (\ref{#1})}
\newcommand{\eqnref}[1]{Eq. (\ref{#1})}
\newcommand{\figref}[1]{Fig. \ref{#1}}
\newcommand{\secref}[1]{Sec. \ref{#1}}
\newcommand{\bea}{\begin{eqnarray}}
\newcommand{\eea}{\end{eqnarray}}

\def\lambdabar{\protect\@lambdabar}
\def\@lambdabar{%
\relax \bgroup
\def\@tempa{\hbox{\raise.73\ht0
\hbox to0pt{\kern.2\wd0\vrule width.7\wd0
height.1pt depth.1pt\hss}\box0}}%
\mathchoice{\setbox0\hbox{$\displaystyle\lambda$}\@tempa}%
{\setbox0\hbox{$\textstyle\lambda$}\@tempa}%
{\setbox0\hbox{$\scriptstyle\lambda$}\@tempa}%
{\setbox0\hbox{$\scriptscriptstyle\lambda$}\@tempa}%
\egroup }

\begin{document}

\title{Letter of Intent: Towards a Vacuum Birefringence Experiment at the Helmholtz International Beamline for Extreme Fields}

\subtitle{(BIREF@HIBEF Collaboration)}

\author[1]{N.~Ahmadiniaz}
\author[1]{C.~B\"ahtz}
\author[2]{A.~Benediktovitch}
\author[2]{C.~B\"omer}
\author[2]{L.~Bocklage}
\author[1,3]{T.~E.~Cowan}
\author[4]{J.~Edwards}
\author[1]{S.~Evans}
\author[1]{S.~Franchino~Vi\~nas}
\author[5,6]{H.~Gies}
\author[7]{S.~G\"ode}
\author[5]{J.~G\"ors}
\author[1]{J.~Grenzer}
\author[1,8]{U.~Hernandez~Acosta}
\author[4]{T.~Heinzl}
\author[6]{P.~Hilz}
\author[5]{W.~Hippler}
\author[1]{L.~G.~Huang}
\author[7]{O.~Humphries}
\author[5,6,9]{F.~Karbstein}
\author[6]{P.~Khademi}
\author[4]{B.~King}
\author[1]{T.~Kluge}
\author[1]{C.~Kohlf\"urst}
\author[2]{D.~Krebs}
\author[1]{A.~Laso-Garc\'{\i}a}
\author[5]{R.~L\"otzsch}
\author[10]{A.~J.~Macleod}
\author[6,9]{B.~Marx-Glowna}
\author[5,6]{E.~A.~Mosman}
\author[7]{M.~Nakatsutsumi}
\author[5]{G.~G.~Paulus}
\author[7]{S.~V.~Rahul}
\author[7]{L.~Randolph}
\author[2,5,6,9]{R.~R\"ohlsberger}
\author[2]{N.~Rohringer}
\author[6]{A.~S\"avert}
\author[6]{S.~Sadashivaiah}
\author[1,3]{R.~Sauerbrey}
\author[1]{H.-P.~Schlenvoigt}
\author[1,3]{S.~M.~Schmidt}
\author[1,3]{U.~Schramm}
\author[1,3]{R.~Sch\"utzhold}
\author[7]{J.-P.~Schwinkendorf}
\author[5,6,9]{D.~Seipt}
\author[1]{M.~\u{S}m\'{\i}d}
\author[5,6,9]{T.~St\"ohlker}
\author[1]{T.~Toncian}
\author[6]{M.~Valialshchikov}
\author[5]{A.~Wipf}
\author[7]{U.~Zastrau}
\author[5,6,9]{M.~Zepf}

\affil[1]{Helmholtz-Zentrum Dresden-Rossendorf, Bautzner Landstra\ss e 400, 01328 Dresden, Germany}
\affil[2]{Deutsches Elektronen-Synchrotron DESY, Notkestrasse 85, 22607 Hamburg, Germany}
\affil[3]{Technische Universit\"at Dresden, 01062 Dresden, Germany}
\affil[4]{Centre for Mathematical Sciences, University of Plymouth, Plymouth, PL4 8AA, UK}
\affil[5]{Department of Physics and Astronomy, Abbe Center of Photonics, Friedrich-Schiller-Universit\"at Jena, 07743 Jena, Germany}
\affil[6]{Helmholtz Institute Jena, Fr\"obelstieg 3, 07743 Jena, Germany}
\affil[7]{European XFEL GmbH, Holzkoppel 4, 22869 Schenefeld, Germany}
\affil[8]{Center for Advanced Systems Understanding, Untermarkt 20, 02826 G\"orlitz, Germany}
\affil[9]{GSI Helmholtzzentrum f\"ur Schwerionenforschung, Planckstra\ss e 1, 64291 Darmstadt, Germany}
\affil[10]{ELI Beamlines Facility, The Extreme Light Infrastructure ERIC, Doln\'{i} B\v{r}e\v{z}any, Czech Republic}

\abstract{Quantum field theory predicts a nonlinear response of the vacuum to strong electromagnetic fields of macroscopic extent. This fundamental tenet has remained experimentally challenging and is yet to be tested in the laboratory. A particularly distinct signature of the resulting optical activity of the quantum vacuum is vacuum birefringence. This offers an excellent opportunity for a precision test of nonlinear quantum electrodynamics in an uncharted parameter regime. 
Recently, the operation of the high-intensity laser ReLaX provided by the Helmholtz International Beamline for Extreme Fields (HIBEF) has been inaugurated at the High Energy Density (HED) scientific instrument of the European XFEL.
We make the case that this worldwide unique combination of an x-ray free-electron laser and an ultra-intense near-infrared laser together with recent advances in high-precision x-ray polarimetry, refinements of prospective discovery scenarios, and progress in their accurate theoretical modelling have set the stage for performing an actual discovery experiment of quantum vacuum nonlinearity.
}


\maketitle

\clearpage 
\tableofcontents
\bigskip 

\section{Executive Summary}
\label{sec:summay}

This letter sets out the intention to perform a first measurement of vacuum birefringence with real photons as a flagship experiment at the HED-HIBEF instrument. 
Photons from the European X-ray Free Electron Laser (EuXFEL) will be scattered at regions of the quantum vacuum polarised by the optical Relativistic Laser at XFEL (ReLaX). Their polarisation will be measured and compared to predictions from quantum electrodynamics (QED). Counting the number of scattered photons that have flipped or not flipped their polarisation allows experimental determination of the low-energy effective field theory couplings of QED, first calculated over 80 years ago \cite{Euler:1935zz,Euler:1935qgl,Heisenberg:1936nmg}. Measurement of vacuum birefringence with real photons can thus be viewed as a first step on the way to harnessing the nonlinearity of the quantum vacuum.

QED predicts a self-interaction of the electromagnetic field that is mediated by virtual electron-positron pairs (some of the constituents of the `quantum vacuum'). This effect is purely quantum mechanical in nature: in classical electromagnetism, the electromagnetic field obeys the superposition principle. The self-interaction of the electromagnetic field has been observed in the scattering of gamma rays in the Coulomb field of atomic nuclei (Delbr\"uck scattering) \cite{Jarlskog:1973aui,Akhmadaliev:1998zz} and more recently at the ATLAS and CMS detectors in the generation of two real photons in the collision of two Coulomb fields \cite{ATLAS:2017fur,ATLAS:2019azn,CMS:2018erd}. So-called `vacuum polarisation' has also been invoked to describe anomalous polarisation of photons measured from strongly magnetised neutron stars \cite{Mignani:2016fwz}. In addition, indirect evidence for vacuum birefringence in the modulation of pairs created via the linear Breit-Wheeler process in the STAR experiment has been noted \cite{Brandenburg:2022tna}. However, the very small cross-section makes direct probing of virtual electron-positron pairs by colliding and scattering only real photons extremely challenging and has yet to be achieved. 
The high number of photons available in laser beams, and the coherence of their electromagnetic fields over spacetime scales much larger than that typically probed by a virtual pair, suggest that colliding focussed laser pulses would be a suitable way to measure this effect. Because the leading-order process is a four-photon interaction, many signatures suggested to be probed in laser experiments are similar to those from four-wave mixing. Among others, these include: manipulation of the polarisation of intense laser pulses, frequency-shifting effects \cite{McKenna:1963,Varfolomeev:1966,Lundstrom:2005za,Lundin:2006wu}, vacuum diffraction \cite{DiPiazza:2006pr,King:2010kvw,Tommasini:2010fb} and vacuum self-focussing \cite{Marklund:2006my,Kharzeev:2006wg}. 
At the same time, any detection in laser beam collisions is challenged by separating the signal of a modified polarisation, momentum or energy in the scattered photons from the large background of the laser fields.

The first direct measurement of vacuum birefringence has been envisioned as a flagship experiment for HED-HIBEF from its initial inception in 2011. HED-HIBEF combines x-rays with an optical pump beam thereby considerably increasing the cross-section, which scales with the centre-of-mass energy to the sixth power, compared to all-optical set-ups. High-precision x-ray polarimetry has been developed over the last decade so that it is now possible to generate a beam of x-rays that are polarised in the same state to a degree of better than one in one hundred billion \cite{Marx:NJP22}. This allows to substantially reduce the background for those x-ray photons that scatter into the `flipped' polarisation mode as a signal of vacuum birefringence.
At the same time the theoretical tools to make quantitatively accurate predictions of quantum vacuum signals in experimentally realistic laser fields have been advancing; see the recent review~\cite{Fedotov:2022ely} and references therein. In light of these developments, we detail several experimental scenarios that can be realised by combining the EuXFEL with the optical laser ReLaX to measure the birefringence of the vacuum with real photons. One of these scenarios features the `dark field' method of blocking part of the XFEL before it is focussed and collides with the optical beam so that in the detector plane in the shadow, there is a region of very few background XFEL photons, which is suitable for detecting a signal \cite{Karbstein:2022uwf}. The actual suppression that can be achieved with the EuXFEL beam will be determined when HED-HIBEF uses its priority access to measure background rates in 2024. In this letter we give details for the experimental implementation of this scenario. Finally, testing QED in an uncharted parameter regime can also constrain the parameter space beyond the Standard Model physics, for instance, that of weakly interacting particles with a small mass \cite{baker13}. In the Summary, the potential of HED-HIBEF experiments to search for new degrees of freedom is discussed.

The fundamental physics prediction of vacuum birefringence represents a prime example of a fascinating and {\it a priori counter-intuitive} phenomenon that can happen when a seemingly trivial state (``the vacuum'') is subjected to extreme conditions (``ultra-intense electromagnetic fields''). Its detection with state-of-the-art technology constitutes a formidable challenge and is at the edge of what is currently possible. The BIREF@HIBEF collaboration brings together experts in strong-field QED, x-ray optics, high-intensity lasers, and laser-plasma physics and thus adopts an interdisciplinary approach to meeting this challenge and performing a discovery experiment.

\section{Introduction}
\label{sec:intro}

The quantum vacuum, i.e.\ the ground state of a quantum field theory such as QED, behaves as a nonlinear, polarisable medium in reaction to strong electromagnetic fields. An electromagnetic wave probing the polarised vacuum may in turn change its polarisation state as a result of \emph{vacuum birefringence}. Microscopically, this is caused by the possibility of \emph{light-by-light scattering} in QED. 

\subsection{History and Status}

\begin{figure}[h!]
\begin{center}
\includegraphics[scale=0.5]{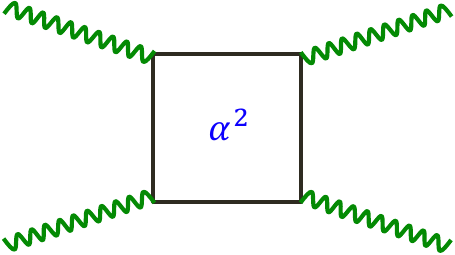}
\caption{\label{fig:LBLS}$O(\alpha^2)$ Feynman diagram for the LbL scattering amplitude implying a 4-photon self-interaction.}
\end{center}
\end{figure}

The idea of light-by-light (LbL) scattering by now has a venerable history. According to Scharnhorst's thorough account \cite{Scharnhorst:2017wzh}, its absence in the classical theory has already been stated by Kepler. In modern terms, this is explained by the linearity of Maxwell's equations of electrodynamics entailing the superposition principle: electromagnetic field solutions can be added at will and remain solutions. In \emph{quantum} electrodynamics, the situation is different. The presence of vacuum polarisation due to virtual pair fluctuations implies the existence of four-photon scattering amplitudes as first pointed out by Halpern \cite{Halpern:1933dya} and Debye (according to Heisenberg \cite{Heisenberg:1934pza}). The presence of this amplitude, represented by the Feynman diagram of Fig.~\ref{fig:LBLS}, implies that the quantum theory becomes nonlinear and photons (self) interact through an effective four-photon vertex. This fact becomes manifest in terms of the celebrated Heisenberg-Euler (HE) Lagrangian, a low-energy effective field theory for QED \cite{Heisenberg:1936nmg} discussed in more detail below.

Heisenberg's students Euler and Kockel did the first calculation of the LbL cross section at low energy \cite{Euler:1935zz}, while Landau's students Akhiezer and Pomeranchuk obtained the high-energy result \cite{Akhiezer:1936vzu}. Due to the four Lorentz indices of the amplitude, the full cross section is somewhat tedious to work out, and it took until 1950 for a full answer to materialise \cite{Karplus:1950zza,Karplus:1950zz} (see also \cite{DeTollis:1964una,DeTollis:1965vna}). Since then, the cross section has become (advanced) textbook material \cite{Akhiezer:1986yqm,Jauch:1976ava, Berestetskii:1982qgu,Itzykson:1980rh,Dittrich:2000zu}. 

As usual in QED, the cross section for LbL scattering may be constrained via dimensional analysis. The basic QED parameters are Planck's constant, $\hbar$, and the speed of light, $c$, which signal the unification of quantum mechanics with special relativity. Henceforth, though, we will choose natural units, $\hbar = c = 1$, unless stated otherwise. The remaining parameters are thus the electron mass and charge, $m$ and $e$, respectively. The latter defines the QED coupling strength, $\alpha = e^2/4\pi = 1/137$, which tells us that QED is (normally) perturbative. We may also form the typical QED length scale, given by the electron Compton wavelength, $\lambdabar_e := 1/m \simeq 3.8 \times 
10^{-13}\,{\rm m}$, and the typical QED electric field magnitude, $E_S := m^2/e\simeq1.3\times10^{18}\,{\rm V}/{\rm m}$, also known as the Sauter-Schwinger limit \cite{Sauter:1931zz,Schwinger:1951nm}. 
Any electric field, when localised within a Compton wavelength, will contain modes above threshold ($q^2 > 4 m^2$) and thus produce pairs with a probability given by a perturbative amplitude (squared) \cite{Itzykson:1980rh}. If the field magnitude exceeds the Sauter-Schwinger limit, non-perturbative, sub-threshold pair production becomes possible.

\begin{figure}[h!]
\begin{center}
\includegraphics[width=9cm]{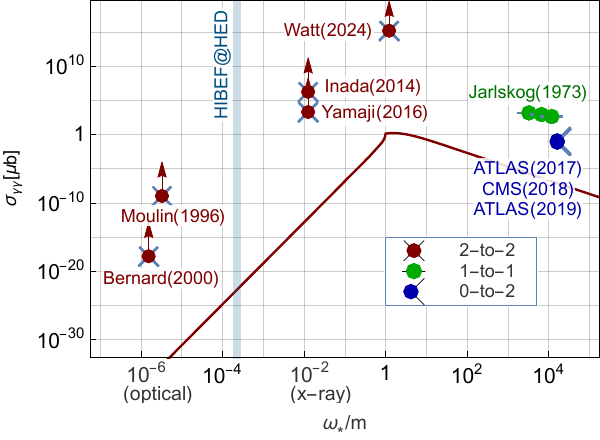}
\caption{\label{fig:LBL.XSECTION}Light-by-light scattering cross section in microbarn ($1\,\upmu{\rm b} = 10^{-30}\,{\rm cm}^2$) as a function of photon energy in the centre-of-mass frame, $\omega_{\ast}$ (calculated from \cite{DeTollis:1964una,DeTollis:1965vna,Costantini:1971cj}). The value of the cross-section probed by some past experiments is indicated by coloured dots `decorated' by `laser lines'. This labelling of the processes (e.g. $2$-to-$2$) refers to the number of real photons in the in and out state (see insert).
The $0$-to-$2$ measurements (blue) were for a diphoton mass $> 5\,\textrm{GeV}$ (CMS, \cite{CMS:2018erd}) or $> 6\,\textrm{GeV}$ (ATLAS, \cite{ATLAS:2017fur,ATLAS:2019azn}).  These are collectively represented on the plot at $\omega_{\ast}/m=\sqrt{s}/2m=5\,\textrm{GeV}$.
The $1$-to-$1$ process of Delbr\"uck scattering off nuclei measured by Jarlskog et al.\ (\cite{Jarlskog:1973aui}, see also \cite{Schumacher:1975kv,Akhmadaliev:1998zz}) is shown in green. An energy range is indicated (in cyan) for HED-HIBEF assuming a near head-on collision of an optical beam with central energy $1.55\,\trm{eV}$ and XFEL beam with energies between $6\,\trm{keV}$ and $12.9\,\trm{keV}$. The red cross-dots represent the $2$-to-$2$ laser experiments by Moulin et al.\ \cite{Moulin:1996vv}, Bernard et al.\ \cite{Bernard:2000ovj} (both all-optical), Inada et al.\ \cite{Inada:2017lop}, Yamaji et al.\ \cite{Yamaji:2016} (both employing an XFEL) and Watt et al.\ \cite{Watt:2024}. 
}
\end{center}
\end{figure}

With the parameters defined, we return to the LbL cross section. At low energies, the dominant energy scale is the electron mass, so the cross section (an area) must be proportional to $1/m^2$. At high energies, masses are irrelevant, and the cross section is inversely proportional to the square of the total energy $\omega_*$ in the centre-of-mass (CM) frame, $\sigma \sim 1/\omega_*^2$. The precise results for the total unpolarised cross section are \cite{Berestetskii:1982qgu}
\begin{alignat}{2}
   \sigma &= \frac{973 \, \alpha^4}{10125 \, \pi}  
   \left( \frac{\omega_*}{m} \right)^6 \frac{1}{m^2}  
   \qquad && (\omega_* \ll m) \; , \label{LBL.LE} \\[5pt] 
   \sigma &= 4.7 \, \alpha^4/\omega_*^2 \qquad && (\omega_* \gg m) 
   \; . \label{SIGMA.HE}
\end{alignat}
The energy dependence of the cross section is depicted in Fig.~\ref{fig:LBL.XSECTION} together with the outcome of some past experiments to be discussed below.

We note that the cross section (\ref{LBL.LE}) is universal in the sense that it is entirely given in terms of QED parameters ($\alpha$ and $m$) and the Lorentz invariant kinematic factor $s = 4 \omega_*^2$, the total energy (squared) in the CM frame. Thus, measuring the cross section is indeed another experimental test of QED. The actual observable is the number $N'$ of scattered photons given by
\begin{eqnarray}
  N' = N n_{\rm L} \Delta z \,  \sigma \; ,
\end{eqnarray}
where $N$ is the number of incoming (probe) photons, $n_{\rm L} =N_{\rm L}/V$ the target photon density and $\Delta z$ the target thickness, i.e.\ the spatial extent of the probed photon distribution. 

At CM energies of order $\omega_* \sim m \sim 0.5\,{\rm MeV}$, the cross section is of the order of $10^{-30}\,{\rm cm}^2$ which, although five orders of magnitude smaller than the Thomson cross section, $\sigma_\mathrm{Th} = (8\pi/3) \alpha^2/m^2$, is not particularly small. However, in this energy regime, available photon fluxes (i.e., $N$ and $N_{\rm L}$) are too small to lead to a measurable number $N'$ of events. At low energies, the situation is reversed. For example, in the optical regime, one has large photon fluxes, but the cross section is exceedingly small, $\sigma \sim 10^{-64}\,{\rm cm}^2$ for $\omega_* = 1.5\,{\rm eV}$. Nevertheless, it has been suggested early on that (intense) lasers may be employed to measure this process \cite{Kroll:1962,McKenna:1963,Harutyunian:1963sao}. The current bound on the cross section from an all-optical scattering experiment is $\sigma < 1.5 \times 10^{-48}\,{\rm cm}^2$ at $\omega_* = 0.8\,{\rm eV}$ \cite{Bernard:2000ovj}, a considerable improvement of the earlier result \cite{Moulin:1996vv}. With the advent of XFELs, the CM energy can be increased by about four orders of magnitude (from $1\,{\rm eV}$ to $10\,{\rm keV}$). Experiments at SACLA (Japan) have produced a bound of $\sigma <  1.9 \times 10^{-23}$ cm$^2$ at $\omega_* = 6.5\,{\rm keV}$ \cite{Yamaji:2016}. At this energy, the QED prediction is $2.5 \times 10^{-43}\,{\rm cm}^2$. So both experiments are off the QED value by about $20$ orders of magnitude, see Table~\ref{tab:LBLS.EXP}. For comparison, we note that the cross section for neutrino electron scattering (for neutrino energies of $10^2\,{\rm eV}$) is about $10^{-58}\,{\rm cm}^2$ and has not been measured yet \cite{Formaggio:2012cpf}.

\begin{table}[h!]
\centering
\begin{tabular}{|l|l|l|l|l|}
\hline
Facility & Experiment &  $\omega_*$ &  Bound & QED value \\
\hline
\hline
LULI & all-optical (two beam) \cite{Moulin:1996vv} & $1.7\;{\rm eV}$ & $\sigma < 9.9 \times 10^{-40}\,{\rm cm}^2$ & $1.6 \times 10^{-64}\;{\rm cm}^2$ \\
LULI & all-optical (three beam) \cite{Bernard:2000ovj} & $0.8\;{\rm eV}$ & $\sigma < 1.5 \times 10^{-48}\;{\rm cm}^2$ & $10^{-66}\;{\rm cm}^2$ \\
SACLA & XFEL + XFEL \cite{Yamaji:2016} & $6.5\,{\rm keV}$ & $\sigma <  1.9 \times 10^{-23}\;{\rm cm}^2$ & $2.5 \times 10^{-43}\;{\rm cm}^2$ \\
HED-HIBEF  & XFEL ($8766$ eV) + optical & $116 \;{\rm eV}$ & & $1.81 \times 10^{-53}\;{\rm cm}^2$ \\
\hline 
\end{tabular}
\caption{\label{tab:LBLS.EXP} Experimental bounds obtained for the LbL scattering cross section and QED predictions (in historical order). The last row refers to the experiment proposed in this LoI which aims to reach the sensitivity for the QED value stated.}
\end{table}

\begin{figure}[h!]
\centering
\begin{subfigure}{.4\textwidth}
  \centering
  \raisebox{0.2\height}{
  \includegraphics[width=.6\linewidth]{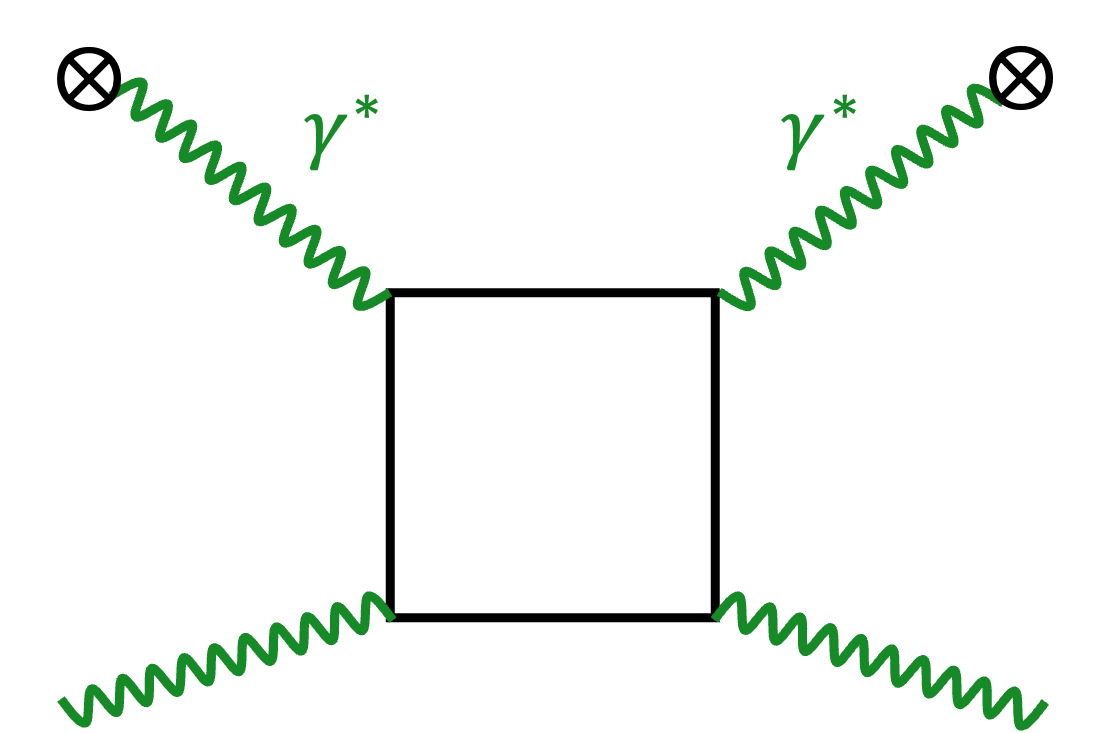}}
  \caption{}
  \label{fig:sub1}
\end{subfigure}%
\begin{subfigure}{.6\textwidth}
  \centering
  \includegraphics[width=.4\linewidth]{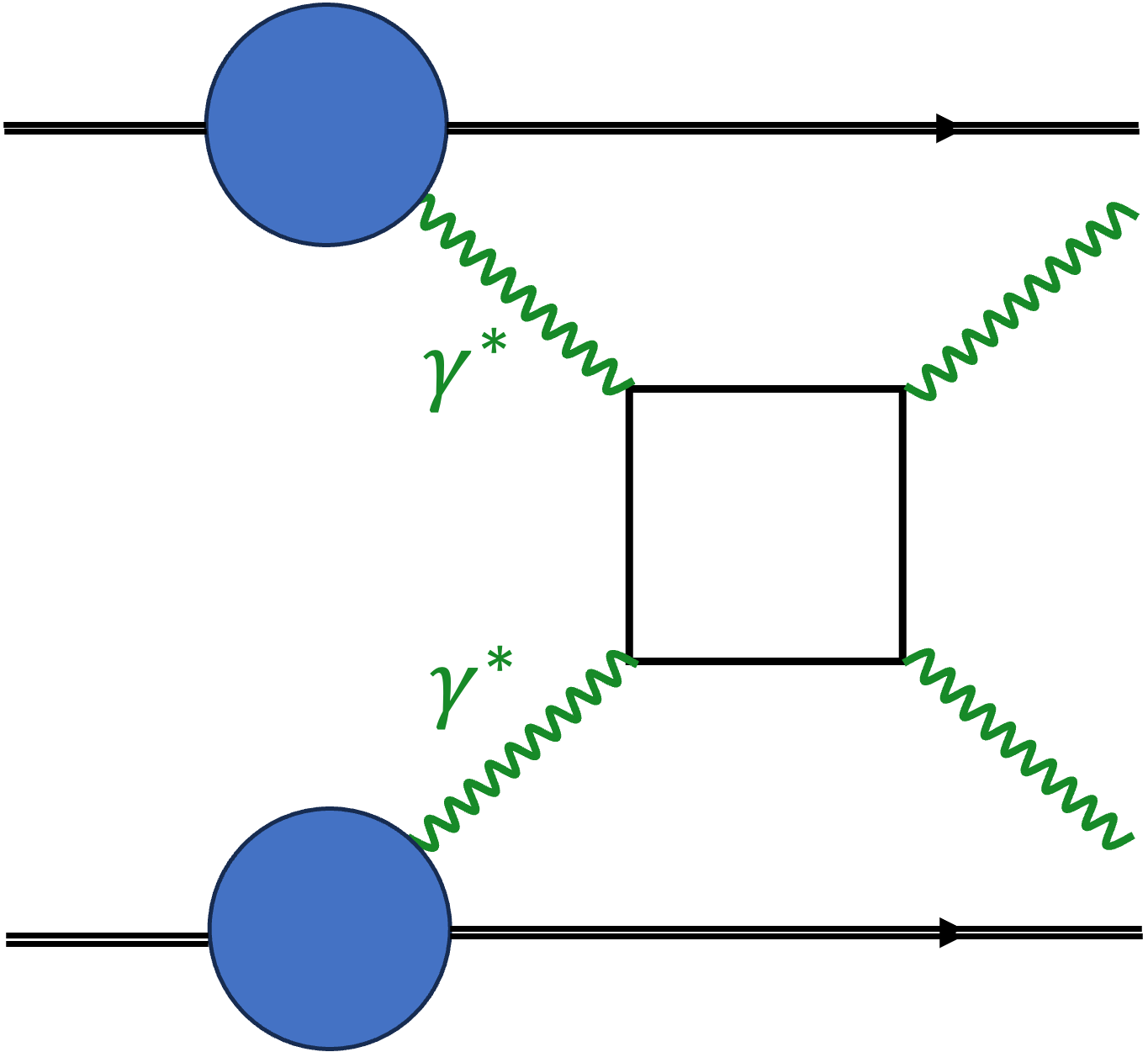}
  \caption{}
  \label{fig:sub2}
\end{subfigure}
\caption{\label{fig:DELBRUECK}Two variants of Delbr\"uck scattering involving two \emph{virtual} photons, $\gamma^*$. (a) Off a (static) Coulomb potential denoted by crosses. (b) Off a Lorentz boosted Coulomb potential in ultra-peripheral heavy-ion collisions.}
\label{fig:test}
\end{figure}

Fig.~\ref{fig:LBL.XSECTION} also shows some experiments at high energy, $\omega_* \gtrsim m$. These have actually observed variants of Delbr\"uck scattering \cite{Delbruck:1933pla} where photons couple to nuclear Coulomb fields. This may be realised by scattering photons off nuclei (charge $Ze$) [Fig.~\ref{fig:DELBRUECK} (a)] or via ultra-peripheral heavy-ion collisions [Fig.~\ref{fig:DELBRUECK} (b)]. In either case, the two intermediate photons are virtual, with their four-momentum $q$ off-shell, $q^2 \ne 0$. Counting vertices and dimensions, the cross section for $\omega_* < m$ can be estimated as \cite{Berestetskii:1982qgu}
\begin{equation} \label{DELBRUECK}
    \sigma \sim Z^4 \alpha^6 \left(\frac{\omega_*}{m}\right)^4 \frac{1}{m^2} \; .
\end{equation}
This yields an increase compared to LbL, recall (\ref{LBL.LE}), if $Z^2 \alpha > \omega_*/m$ which is easily achieved for $\omega_* < m$. In particular, the nuclear charge can be viewed as a \emph{coherent} enhancement factor which is also present at larger energies, $\omega_* > m$ \cite{Berestetskii:1982qgu}. As a result, the Delbr\"uck processes of Fig.~\ref{fig:DELBRUECK} have indeed been measured, both for nuclei at rest (Fig.~\ref{fig:LBL.XSECTION}, green dots, \cite{Jarlskog:1973aui,Schumacher:1975kv,Akhmadaliev:1998zz}) and in heavy-ion collisions (Fig.~\ref{fig:LBL.XSECTION}, blue dots, \cite{ATLAS:2017fur,CMS:2018erd,ATLAS:2019azn}). Fig.~\ref{fig:LBL.XSECTION} clearly shows that the Delbr\"uck cross section exceeds the LbL one -- by at least five orders of magnitude for $\omega_* \gg m$.  For additional details we refer to the overview presented in the introduction of \cite{Ahmadiniaz:2020jgo}.

\subsection{Light-by-light scattering with lasers}

At BIREF@HIBEF an intense optical laser ($\omega_{\rm L} = 1.55\;{\rm eV}$) will be combined with the European XFEL (which can run at different frequencies from about $5$ to $24\;{\rm keV}$, e.g. at $\omega_{\rm X} = 8766\;{\rm eV}$). This implies a CM energy given by the geometric mean of $\omega_* = (\omega_{\rm L} \omega_{\rm X})^{1/2} = 116\;{\rm eV}$. According to (\ref{LBL.LE}), the QED cross section will then be $\sigma = 1.81 \times 10^{-53}\;{\rm cm}^2$. This is still fairly small, but the coherence of the optical laser background drastically enhances the amplitude by a factor proportional to the photon number. 

\begin{figure}[h!]
\begin{center}
\includegraphics[width=0.8\textwidth]{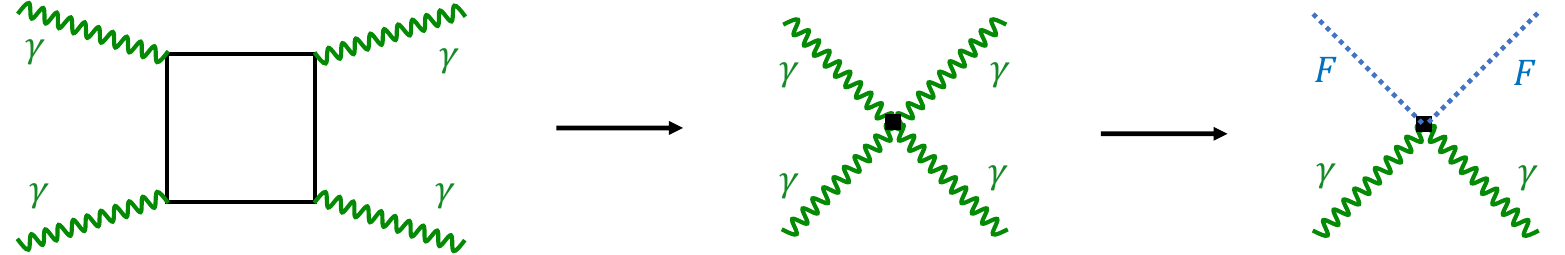}
\caption{\label{fig:LBL-LBLASER}Two-step modification of the QED 4-photon vertex (left) to the Heisenberg-Euler 4-point interaction (centre) and the coupling to an external field scattering (right).}
\end{center}
\end{figure}

The calculation of the relevant cross section thus proceeds in two steps. First, one employs a low-energy approximation by adopting the Heisenberg-Euler effective Lagrangian with a point-like four-photon vertex. Intuitively, the vacuum polarisation loop can no longer be resolved as the effective theory is only valid for distances much larger than the Compton wave length. In a second step, one replaces two of the photon legs in Fig.~\ref{fig:LBLS} by an external electromagnetic field representing the intense laser focus. The two steps are depicted in terms of Feynman diagrams in Fig.~\ref{fig:LBL-LBLASER}.

Scattering off an optical laser background $F$ yields a coherent enhancement factor $F^2$ in the amplitude. This factor is proportional to the photon number $N_{\rm L}$ of the optical laser, hence analogous to the squared nuclear charge, $Z^2$, in Delbr\"uck scattering. Thus, replacing $Z^2$ by $F^2 \sim N_{\rm L}$ in (\ref{DELBRUECK}), the cross section is expected to behave as
\begin{equation} \label{SIGMA.1TO1}
    \sigma \sim \alpha^4 N_{\rm L}^2 \left( \frac{\omega_*}{m}\right)^6 \frac{1}{m^2} \; .
\end{equation}
The dimensionless constant of proportionality can be calculated from the low-energy effective field theory of QED,
\begin{equation}
    \mathcal{L}_\mathrm{eff} = \mathcal{S}
     + \mathcal{L}_{\trm{HE}}
     + c_{22} F_{\mu\nu} \Box F^{\mu\nu} + 
   \ldots \; \qquad \mathcal{L}_{\trm{HE}} = c_{1} \mathcal{S}^2 + c_2 \mathcal{P}^2 \; , \label{eqn:HE1}
\end{equation}
where $\mathcal{L}_{\trm{HE}}$ is the celebrated Heisenberg-Euler (HE) Lagrangian \cite{Heisenberg:1936nmg} to lowest order in field strengths. Its crucial ingredients are the fundamental Lorentz and gauge invariants,
\begin{eqnarray}
  \mathcal{S} &=& - \frac{1}{4} F_{\mu\nu} F^{\mu\nu} = \frac{1}{2} (\vc{E}^2 - \vc{B}^2) \; , \\
  \mathcal{P} &=& - \frac{1}{4} F_{\mu\nu} \tilde{F}^{\mu\nu} = \vc{E} \cdot \vc{B} \; , 
  \label{invs}
\end{eqnarray}
with electromagnetic field strength tensor $F^{\mu\nu} = \partial^\mu A^\nu - \partial^\nu A^\mu$ and its dual, $\tilde{F}^\mu = (1/2) \epsilon^{\mu\nu\rho\sigma}F_{\rho\sigma}$, gauge potential $A^\mu$ and low-energy constants $c_1$, $c_2$ and $c_{22}$. The invariant $\mathcal{S}$ is just the Maxwell Lagrangian while $\mathcal{S}^2$ and $\mathcal{P}^2$ are the leading-order nonlinear corrections first employed in \cite{Euler:1935zz}. The ellipsis represents corrections of higher order in both field strength \cite{Heisenberg:1936nmg} and derivatives \cite{Mamaev:1981dt,Gusynin:1995bc,Gusynin:1998bt,Karbstein:2021obd,Franchino-Vinas:2023wea}. The low-energy constants are 
\begin{equation}
  c_1 = \frac{8 \alpha^2}{45 m^4} \; , \quad
  c_2 = \frac{14\alpha^2}{45 m^4} \; , \quad
  c_{22} = \frac{\alpha}{60\pi m^2} \; ,
  \label{eq:c_i}
\end{equation}
where the powers in $\alpha$ and $m$ follow from vertex counting and dimensional analysis.
These coefficients receive higher loop corrections that are parametrically suppressed with additional powers of $\alpha \ll 1$. They account for higher-order vacuum polarisation effects arising from the interaction of charges in the vacuum polarisation loop cf., e.g., \cite{Gies:2016yaa}.

In this context, we also emphasise that the structure of \eqnref{eqn:HE1} is generic for any nonlinear extension of classical electrodynamics that respects Lorentz covariance, U(1) gauge invariance, and a charge conjugation parity symmetry. Hence, a measurement of the QED predictions in \eqnref{eq:c_i} inherently also implies a restriction of the parameter space of other potential non-linear extensions of classical electrodynamics in and beyond the Standard Model of particle physics.

In QED, the last constant, $c_{22}$, is the weight of the leading-order \emph{linear} vacuum polarisation term and was already calculated by Dirac and Heisenberg in 1934 \cite{Dirac:1934,Heisenberg:1934pza}. Its QED analog, valid for all energies, implies a scale dependence, hence `running', of the electric charge which decreases with distance. Intuitively, this corresponds to charge screening caused by the virtual pair dipoles in the vacuum. This implies in turn that the vacuum can be viewed as a polarisable medium with both linear and nonlinear response to an external field. The associated response functions are given by the second derivatives of the effective Lagrangian with respect to $F$ (or $\vc{E}$ and $\vc{B}$). Focussing on the nonlinear case we can define macroscopic fields $\vc{D}$ and $\vc{H}$ as the derivatives of the Lagrangian $\mathcal{L}_\mathrm{eff} = \mathcal{S} + \mathcal{L}_\mathrm{HE}$,
\begin{eqnarray}
   D_i &=& \frac{\partial \mathcal{L}_\mathrm{eff}}{\partial E_i} \equiv \varepsilon_{ij} E_j  \; , \\
   H_i &=& - \frac{\partial \mathcal{L}_\mathrm{eff}}{\partial B_i} \equiv \mu_{ij}^{-1} B_j  \; ,
\end{eqnarray}
where the (static) nonlinear response functions are the tensors \cite{Baier:1967zzc}
\begin{eqnarray}
  \varepsilon_{ij} &=& (1 + 2 c_1 \mathcal{S}) \, \delta_{ij} + 2 c_2 \, B_i B_j \; , \label{EPSILON}\\
  \mu_{ij}^{-1} &=& (1 + 2 c_1 \mathcal{S}) \,  \delta_{ij} - 2 c_2 \, E_i E_j \; . \label{MU.INV}
\end{eqnarray}
These correspond to the dielectric and permeability 
tensors, respectively, and thus explicitly show the electric and magnetic response of the vacuum to external fields. In the words of Weisskopf \cite{Weisskopf:1936hya}: `\textit{When passing through electromagnetic fields, light will behave as if the vacuum had acquired a dielectric constant different from unity due to the influence of the fields.}'

From conventional optics, it is known that media with dielectric constant $\varepsilon$ and permeability $\mu$ have optical properties characterised by an index of refraction, $n \equiv \sqrt{\varepsilon \mu}$. In view of (\ref{EPSILON}) and (\ref{MU.INV}) it must thus be possible to ascribe a nontrivial refractive index to the vacuum. This can indeed be done by studying the eigenvalues of the dielectric and permeability tensors, $\varepsilon_{ij}$ and $\mu_{ij}$ \cite{Klein:1964,Dittrich:2000zu,Koch:2005}. One finds that the vacuum is characterised by \emph{two} nontrivial refractive indices, $n_1$ and $n_2$, so that the vacuum acts like a \emph{birefringent} medium.

\subsection{Vacuum Birefringence}

A web search for the term `vacuum birefringence' dates its first appearance to the year 1964 when it appeared in the title of a paper by Klein and Nigam \cite{Klein:1964} who employed the tensors (\ref{EPSILON}) and (\ref{MU.INV}) stemming from the low-energy HE Lagrangian.
However, the vacuum refractive indices were first calculated (for arbitrary photon energies!) by Toll in his unpublished PhD thesis  \cite{Toll:1952} supervised by Wheeler. Other early and important contributions have been made by Erber \cite{Erber:1961} (preceding Klein and Nigam!), Baier and Breitenlohner \cite{Baier:1967zzc} and Narozhny \cite{Narozhny:1969}.
The utilisation of lasers has first been suggested and analysed in \cite{Aleksandrov:1985}. The HED-HIBEF experiment is based on the optimal scenario of combining an XFEL probe beam with an intense optical laser background. This was first suggested in \cite{Heinzl:2006xc}. 

To avoid confusion we recall that the term `vacuum birefringence' refers to the refractive or dispersive properties of the vacuum (encoded in the real part of the refractive indices), while the term `vacuum dichroism' describes its absorptive properties (encoded in the imaginary part of the refractive indices). In other words, dichroism amounts to a direction dependent absorption of probe photons, hence a reduction of probe intensity \cite{Klein:1964zza,Heyl:1997hr}.

A quick way to derive the magnitude of the birefringence effect is to employ the differential cross section in the CM frame which for a $2 \to 2$ process (with all masses equal) has the universal form
\begin{equation}
    \frac{d\sigma}{d\Omega} = \frac{1}{64\pi^2} \frac{1}{s} \, |\mathfrak{M}|^2 \equiv |f(\vc{l}, \vc{l}', \theta)|^2 \; .
\end{equation}
Here $\mathfrak{M}$ is the invariant amplitude depending on photon polarisations and momenta while $f$ is the scattering amplitude for momentum transfer, $\vc{l} \to \vc{l}'$, and scattering angle $\theta$. Microscopically, vacuum birefringence corresponds to forward scattering ($\vc{l} =\vc{l}'$, $\theta = 0$) with a polarisation flip of a probe photon passing through an intense laser background. We thus assume the involvement of two laser photons with the same polarisations and four vectors, while the probe polarisation flips, $\varepsilon \to \varepsilon'$ with $\varepsilon \cdot \varepsilon' = 0$. Choosing the optimal scenario of a $45^\circ$ angle between probe and laser polarisations, one may use the textbook formulae in \cite{Akhiezer:1986yqm} to find the forward-flip amplitude $\mathfrak{M}(0)$ and hence the forward scattering amplitude, 
\begin{equation}
    f(\vc{l}, \vc{l}, 0) \equiv f(0) = \frac{\mathfrak{M}(0)}{8\pi \sqrt{s}} = \frac{4 \alpha^2}{15 \pi} \left( \frac{\omega_*}{m}\right)^3 \frac{1}{m} \; .
\end{equation}
Another textbook formula (see e.g.\ \cite{Newton:1982qc}, Ch.~1.5) relates the index of refraction to the forward scattering amplitude. The flip amplitude, in particular, defines the \emph{difference} of the vacuum refractive indices, 
\begin{equation} \label{DELTA.N}
    \Delta n = \frac{2\pi n_{\rm L}}{\omega_*^2} f(0) = 
    \frac{8 \alpha^2}{15} \frac{I_{\rm L}}{m^4} \; .
\end{equation}
Here we have used that the laser photon density, $n_{\rm L} = N_{\rm L}/V = I_{\rm L}/\omega_*$, is proportional to intensity, $I_{\rm L}$, measured in the CM system. This result is consistent with the intensity scaling that led to (\ref{SIGMA.1TO1}) and the calculation \cite{Narozhny:1969} of the refractive indices in terms of the vacuum polarisation tensor (the covariant unification of dielectric and permeability tensors),
\begin{equation}
    n_i = 1 + 4 c_i I_{\rm L} \quad \mbox{whence} \quad \Delta n = 4 (c_2 - c_1) I_{\rm L} \; .
\end{equation}
This may be written in a more covariant way as follows. One introduces the probe 4-vector $k = \omega \ell$ with frequency $\omega$ and a dimensionless null-vector $\ell$, $\ell^2 = 0$. One then forms the `null-energy projection' \cite{Shore:2007um}, $T_{\ell\ell} \equiv \ell_\mu T^{\mu\nu} \ell_\nu$, of the Maxwell energy-momentum tensor, $T^{\mu\nu} = F^{\mu\alpha} F_\alpha^{\;\;\nu} - g^{\mu\nu} \mathcal{S}$, the traceless part of the squared field strength. For any field configuration, $F^{\mu\nu}$, one can thus write
\begin{eqnarray} \label{DELTA.C}
    \frac{\Delta n}{T_{\ell\ell}} = c_2 - c_1 = \frac{2\alpha^2}{15 m^4} \; ,
\end{eqnarray}
which obviously just measures the difference of the leading-order low-energy constants in the HE Lagrangian. 
The difference $\Delta n$ in refractive indices induces an ellipticity $\delta \equiv k_{\rm X} z \Delta n/2 $ for a linearly polarised probe beam of wave number $k_{\rm X}$ traversing the polarised vacuum across a distance $z$. The experimental signature is then the flip probability,
\begin{equation}
    \frac{N'}{N} = \delta^2 = \frac{16 \alpha^4}{225} \frac{I_{\rm L}^2}{m^8} \left(\frac{z}{\lambdabar_{\rm X}}\right)^2 = 
    \frac{4 \alpha^2}{225} \left(\frac{I_{\rm L}}{I_{\rm S}}\right)^2 \left(\frac{z}{\lambda_{\rm X}}\right)^2\; ,
    \label{eq:deltasquared}
\end{equation}
with the Sauter-Schwinger intensity $I_{\rm S} = E_{\rm S}^2 = m^4/4\pi\alpha = 4.7 \times 10^{29}\,{\rm W}/{\rm cm}^2$. Eq.~(\ref{eq:deltasquared}) shows explicitly that an optimal scenario will maximise the target intensity $I_L$ and its spatial extent, $z$, while simultaneously minimising the (reduced) probe wave length, $\lambdabar_{\rm X} \equiv \lambda_{\rm X}/2\pi$ \cite{Heinzl:2006xc}. At HED-HIBEF this is realised by combining a high-intensity optical laser with an XFEL. A rough estimate with $I_{\rm L} = 10^{21}\,{\rm W}/{\rm cm}^2$, $z= 10 \;\upmu\trm{m}$ and $\lambdabar_{\rm X} = 0.02\,{\rm nm}$ (for $\omega_{\rm X} = 10\,{\rm keV}$) yields $N'/N \approx 10^{-12}$. Obviously, such a minuscule flip probability demands a very high accuracy of the required polarisation measurements. Indeed, the original suggestion to employ XFEL beams as a probe \cite{Heinzl:2006xc} has been an incentive to improve the polarisation purity of x-rays by several orders of magnitudes to a current record of $10^{-11}$ \cite{Schulze:2022}. A topical review of the technical difficulties involved and how they are being addressed is given in \cite{Yu:2023}. 

\subsection{Previous Experiments}

\begin{figure}[h!]
\begin{center}
\includegraphics[width=0.7\textwidth]{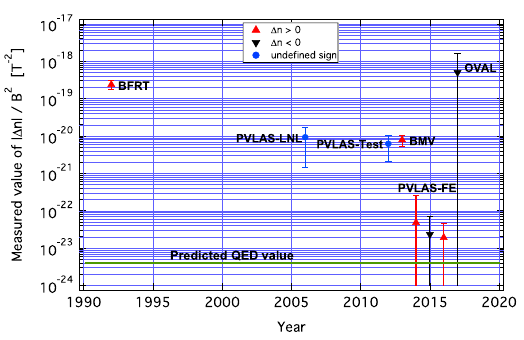}
\caption{\label{fig:PVLAS}Historical evolution of the results for vacuum birefringence experiments employing static magnetic fields  (reproduced from \cite{Ejlli:2020yhk} where more details can be found.) The green horizontal line represents $k_\mathrm{CMV} \equiv c_2 - c_1$.}
\end{center}
\end{figure}

As early as 1929, Watson \cite{Watson:1929ssm} tried to measure a vacuum refractive index $n$ depending linearly on a static magnetic field, $B$. He used an interferometer to measure a small induced frequency shift, but did not find any effect and thus produced the upper bound of $(n-1)/B < 4 \times 10^{-7} \,{\rm T}^{-1}$ (in modern notation). In 1960 Jones \cite{Jones:1960} measured the velocity of light in a magnetic field to high precision and found no deviation from its vacuum value. A year later, Erber \cite{Erber:1961} discussed a number of settings (including Watson's) that might allow measurement of a \emph{quadratic} dependence of the vacuum refractive indices, hence a Cotton-Mouton effect \cite{Cotton-Mouton:1905a,Cotton-Mouton:1905b}, as predicted by QED. Klein and Nigam \cite{Klein:1964} suggested using a static electric field inside a plane capacitor but ruled the birefringence effect to be way too small. In 1979 Iacopini and Zavattini pointed out that one may use strong static magnetic fields and optimise the geometric factor $(z/\lambdabar)^2$. To this end, one should propagate a stable optical laser beam through a magnetic cavity and increase the optical path length $z$ through multiple reflections. This idea has been realised in the PVLAS experiment, which recently celebrated its 20th anniversary \cite{Ejlli:2020yhk}. There are also a number of competing or complementary experiments for which an overview may be found in \cite{Battesti:2018bgc} together with an extensive list of references. Fig.~\ref{fig:PVLAS}, reproduced from \cite{Ejlli:2020yhk}, presents the historical evolution of the experimental results, the most recent of which are still about an order of magnitude above the QED prediction. (The error bars represent an uncertainty of one sigma.) The current best value has been reported by the PVLAS-FE experiment as 
\begin{equation}
    \frac{\Delta n^{(\mathrm{PVLAS-FE})}}{B^2} = (+19 \pm 27) \times 10^{-24} \,{\rm T}^{-2} \; ,
\end{equation}
where $B$ denotes the external \emph{static} magnetic field being probed. To compare with the QED result, we note that for a static magnetic field probed at the right angle, $T_{\ell\ell} = B^2$. Introducing the magnetic Sauter-Schwinger field strength, $B_{\rm S} = m^2/e = 4.41 \times 10^9\,{\rm T}$, the general result (\ref{DELTA.C}) leads to
\begin{equation}
    k_\mathrm{CMV} \equiv \frac{\Delta n^{\mathrm{stat}}}{B^2} = c_2 - c_1 = \frac{2\alpha^2}{15 m^4} = \frac{\alpha}{30 \pi} \frac{1}{B_{\rm S}^2} = 4 \times 10^{-24}\,{\rm T}^{-2} 
    \; .
\end{equation}
This constant, containing only the basic QED parameters, has been called the Cotton-Mouton constant of the vacuum (hence the acronym CMV) in \cite{Battesti:2018bgc}. The same value is obtained for the HED-HIBEF scenario of probing an optical laser background, cf.\ (\ref{DELTA.N}) and (\ref{DELTA.C}), as $T_{\ell\ell} = 4F^2 = 4I_{\rm L}$ (assuming a head-on collision). Hence,  
\begin{equation}
  \frac{\Delta n^{\mathrm{CF}}}{4F^2} = c_2 - c_1 \; ,
\end{equation}
where CF stands for `crossed field' with $F=E=B$ and $\vc{E} \cdot \vc{B} = 0$. 

Strong-field vacuum polarisation effects also play an important role in astrophysics and cosmology. For instance, extreme field magnitudes can be found in the magnetosphere of neutron stars and magnetars. Several years ago it was claimed that measurements of the polarisation degree of such neutron stars provide evidence for an active role of vacuum birefringence \cite{Mignani:2016fwz}. The subsequent scientific debate \cite{Capparelli:2017mlv} highlights the necessity of a controlled laboratory verification of the effect which in turn should provide input for improved models of neutron star environments. 

\begin{figure}
\centering
\begin{subfigure}{.4\textwidth}
  \centering
  \raisebox{0.15\height}{
\includegraphics[width=.7\linewidth]{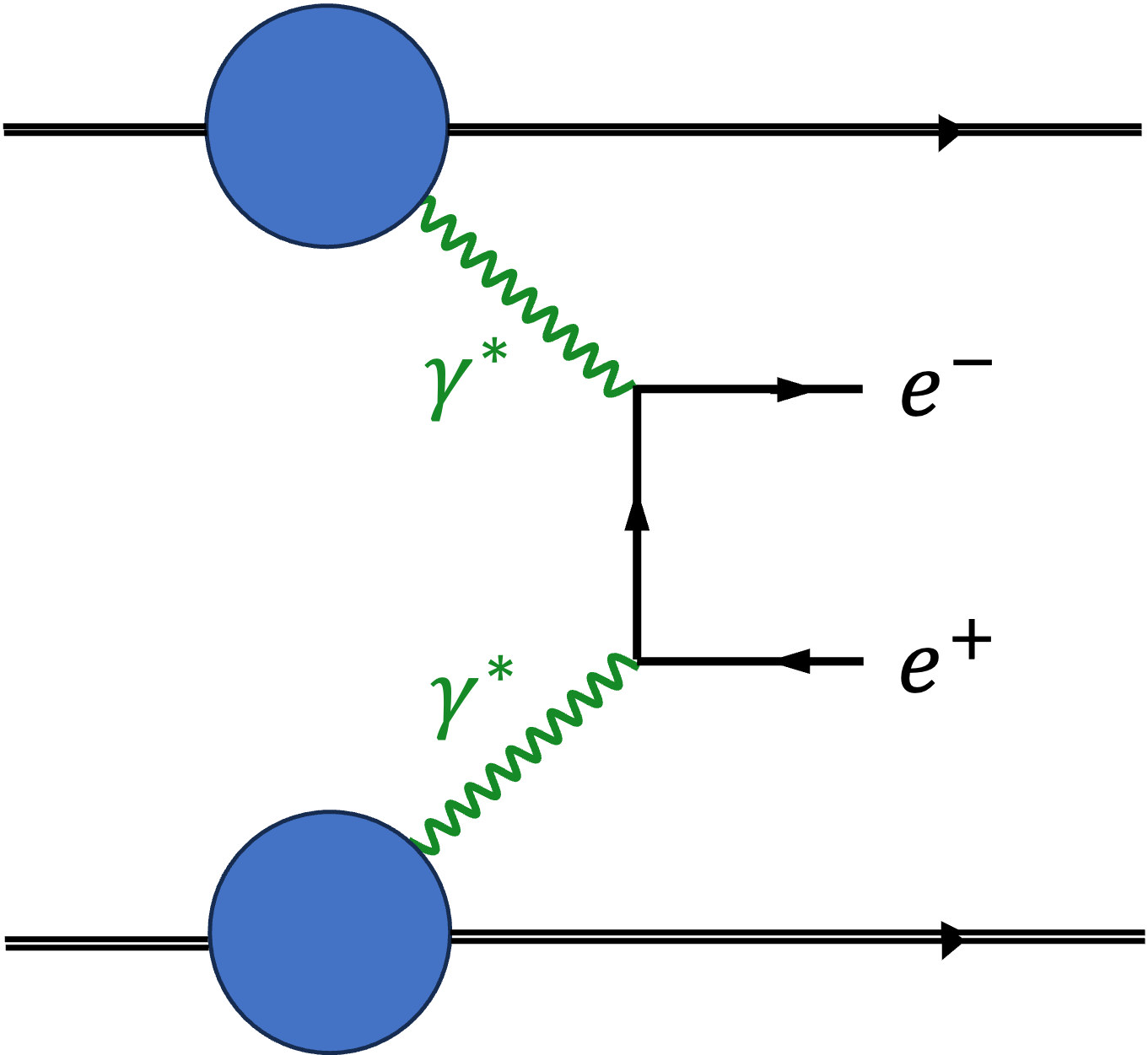}}
  \caption{}
\end{subfigure}%
\begin{subfigure}{.6\textwidth}
  \centering
  \includegraphics[width=.8\linewidth]{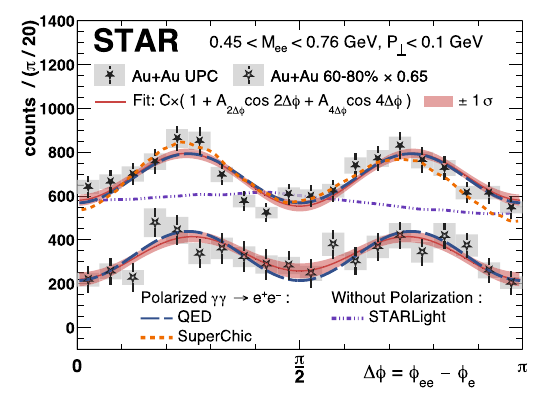}
  \caption{}
\end{subfigure}
\caption{\label{fig:STAR}The STAR experiment. (a) Landau-Lifshitz process. (b) Modulation signalling vacuum birefringence by means of the optical theorem (reproduced from \cite{Brandenburg:2022tna}).}
\end{figure}

Very recently the STAR collaboration has reported the observation of (linear) Breit-Wheeler pair production \cite{STAR:2019wlg}, i.e.\ the process $\gamma + \gamma \to e^+ + e^-$. As this was realised in ultra-peripheral heavy-ion collisions, the pair production is actually proceeding via the Landau-Lifshitz process \cite{Landau:1934} at low photon virtualities, see Fig.~\ref{fig:STAR} (a). Via the optical theorem, this process is related to Delbr\"uck scattering as observed by the ATLAS collaboration -- upon 'cutting' the fermion loop in Fig.~\ref{fig:DELBRUECK} (b). It has been shown that polarisation flips induced by this diagram lead to modulations in the angular pair spectra as shown in Fig.~\ref{fig:STAR} (b) reproduced from the recent review \cite{Brandenburg:2022tna}. The authors of \cite{Brandenburg:2022tna} have interpreted this as an indirect signal for vacuum birefringence.

\section{Prospective Scenarios}
\label{sec:scenarios}
\begin{figure}[h!!]
    \centering
    \includegraphics[width=0.65\linewidth]{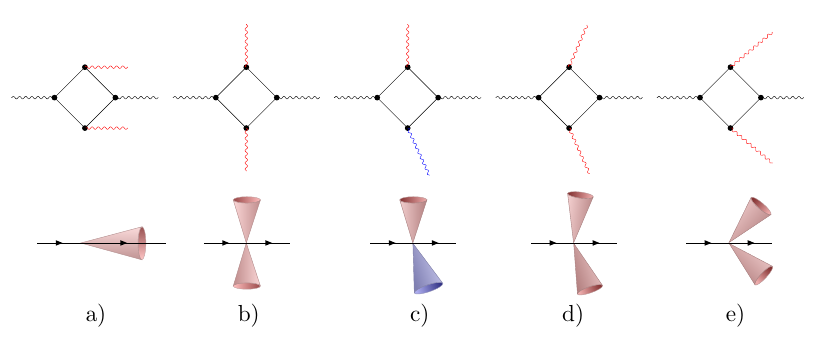}
    \caption{Overview of different collision geometries: (a) is the conventional head-on two-beam scenario; (b)-(e) are three-beam scenarios where the XFEL is collided with two optical lasers. In (c) one of the optical beams is frequency-doubled \cite{Ahmadiniaz:2022nrv}.}
    \label{fig:ralfFigs}
\end{figure}

As detailed in the previous section, real light-by-light scattering has a very small cross-section. Any prospective scenario must be able to deliver a signal that can be distinguished from the noise of the background photons. 
At the moment, three different proposals -- and viable combinations thereof -- are available that constitute prospective routes toward a first measurement of the effect at HED-HIBEF.
All these proposals are intimately related and look very similar at 
first glance. However, they differ decisively in the details and thus come with different experimental challenges.
Two of these are {\it two-beam scenarios} using the collision of the XFEL with the intense ReLaX beam, and the third one summarises {\it three-beam scenarios} requiring the XFEL to collide with two intense laser beams, generated by splitting the original ReLaX beam into two beams. (See Fig.~\ref{fig:ralfFigs} for an overview of different collision geometries.)
Focussing only on the essentials (see Secs.~\ref{subsec:2beams}-\ref{subsec:3beams} below for a detailed discussion of each of these scenarios), the {\bf two-beam scenarios} are:
\begin{itemize}
 \item[(1)] The {\it conventional vacuum birefringence scenario} \cite{Heinzl:2006xc,Schlenvoigt:2016jrd} that envisions the collision of the two beams in a counter propagation geometry. It uses exclusively the polarisation flip of x-ray probe photons. This small quantum vacuum signal needs to be separated from the large background of probe photons traversing the
 high-intensity pump focus without interaction.
 A key parameter for the measurement of this effect is the quality of the employed polarimeter which is typically quantified in terms of its {\it polarisation purity}~$\cal P$.  This represents the fraction of background photons registered in the ideally empty, perpendicularly polarised mode \cite{PhysRevLett.110.254801}. For the XFEL and intense laser parameters attainable at HED-HIBEF the signal is found to be background dominated \cite{Schulze:2022}.
 However, by an appropriate choice of the beam waists, the divergence of the signal can be made wider than that of the probe, such that angular cuts can be used to improve the signal-to-background ratio at the expense of reduced absolute photon numbers \cite{King:2010kvw,King:2010nka,King:2012aw,Karbstein:2015xra}.
  \item[(2)] A {\it dark-field approach} \cite{Karbstein:2022uwf} based on modifying the probe beam with a well-defined beam stop such as to exhibit a shadow in both the converging and expanding beam while retaining a peaked focus profile. This should allow access to both the parallel and perpendicular polarised components of the nonlinear vacuum response scattered into the shadow where the background is significantly reduced.
  In this scenario, the key parameter is the quality of the shadow which can be quantified in terms of the unwanted background measured within the shadow.
  It remains to be shown that a sufficiently good {\it shadow quality} $\cal S$ ensuring a signal-to-background ratio above unity can be realised in experiments for the XFEL and intense laser parameters available at HED-HIBEF.
  However, the outcome of an elementary proof-of-concept experiment at an x-ray tube \cite{Karbstein:2022uwf} and the results of numerical diffraction simulations performed by our collaboration look sufficiently promising that this scenario is the one we have selected to first explore experimentally; cf. also Sec.~\ref{sec:experiment} below.
\end{itemize}
Finally, the {\bf three-beam scenarios} aim at verifying quantum vacuum nonlinearity in
\begin{itemize}
 \item[(3)] {\it four-wave-mixing processes} \cite{Lundstrom:2005za,Lundin:2006wu,Gies:2017ezf,King:2018wtn,Aboushelbaya:2019ncg,Ahmadiniaz:2022nrv,Berezin:2024fxt} where the XFEL is collided with two intense optical laser beams that are derived from the same source and are focussed to the same spot.
Since the optical beams have the same frequency and propagate in different directions, this enables a new kind of quasi-elastic scattering signal at the XFEL photon frequency, the generation of which involves the absorption of a field quantum from one of the intense beams and emission into the other.
Due to the associated finite momentum transfer, this signal is scattered into a well-defined direction away from the forward direction of the XFEL.
Moreover, three-beam setups can also induce sizeable signals characterised by a frequency shift. While an enlarged phase space seems beneficial, it remains to be shown that the additional challenges coming with the experimental control of three laser beams can be mastered in a way such as to benefit from the additional signal photon channels facilitated by three-beam scenarios. 
\end{itemize}

For the experimental implementation of a given scenario, the central interest is in the number of x-ray signal photons that can be {\it discerned} from the typically large background of the EuXFEL beam. Only these constitute a signature of quantum vacuum nonlinearity that is {\it detectable} in an experiment. 
To address this theoretically, we model the near-infrared (NIR) high-intensity and XFEL beams driving the nonlinear quantum vacuum signals as paraxial solutions of the wave equation, supplemented with a Gaussian pulse envelope. In general, the electric field of a paraxial beam can be expanded as $\mbf{E} = \mbf{E}^{(0)}(\varsigma) + \varsigma\, \mbf{E}^{(1)}(\varsigma) + \cdots$, i.e., in powers of the small parameter $\varsigma = w_{0}/z_{\rm R}$, where $w_{0}$ is the beam waist and $z_{\rm R} = \omega w_{0}^{2}/2$ is the Rayleigh length of the fundamental Gaussian beam solution; $\omega=2\pi/\lambda$ is the oscillation frequency of a beam with central wavelength $\lambda$.
Throughout this work, we truncate the paraxial expansion at leading order and use $\mbf{E} \approx \mbf{E}^{(0)}(\varsigma)$.
Apart from the probe beam featuring a central shadow in the dark-field scenario, the description of which requires the superposition of several Laguerre-Gaussian or Hermite-Gaussian modes, we model all laser fields as linearly polarised fundamental Gaussian beams.
For a beam propagating in positive $\rm z$ direction, this implies the following electric field profile,
\begin{equation}
    E=E_0\,\exp\left\{-\left(\frac{t-z}{\tau/2}\right)^2-\left(\frac{r}{w}\right)^2\right\}\cos\left\{\omega(t-z)-\frac{z}{z_{\rm R}}\left(\frac{r}{w}\right)^2+\arctan\frac{z}{z_{\rm R}}\right\}\,,
    \label{eqn:parax1} 
\end{equation}
where $E_0$ is the field amplitude, $\tau$ is the phase pulse duration, $r=\sqrt{x^2+y^2}$ and $w=w_0\sqrt{1+(z/z_{\rm R})^2}$.
This, in particular, implies that the peak intensity $I_{\trm{peak}}$ of the optical laser pulse is related to the pulse energy $W$ via $I_{\rm peak} = 8 \sqrt{2/\pi}\,W/(\pi w_{0}^{2} \tau) = 2\times 10^{23}\,[{\rm Wcm}^{-2}] W[{\rm J}] (w_{0}[\upmu{\rm m}])^{-2}(\tau[{\rm fs}])^{-1}$. The focal width and pulse duration are related to the full-width-at-half-maximum (FWHM) parameters via: $w_{0} = 0.85\,w_{\rm FWHM}$ and $\tau = 1.7 \, \tau_{\rm FWHM}$. Standard parameters to be available at HED-HIBEF are given in Tab.~\ref{tab:params1}. 

\begin{table}[h!!]
\begin{adjustbox}{center}
    \begin{tabular}{ | c | c | }
        \hline
        \multicolumn{2}{| c |}{\bf European XFEL}\\
        \hline
        \hline
        \multirow{4}{*}{$N$} & $2\times10^{11}$\,@\,$\omega=8\ldots10\,{\rm keV}$ (self-seeded)\\
         & $1\times10^{11}$\,@\,$\omega=12\ldots13\,{\rm keV}$ (self-seeded)\\
         & $1\times10^{12}$\,@\,$\omega=8\ldots10\,{\rm keV}$ (SASE)\\
         & $5\times10^{11}$\,@\,$\omega=12\ldots13\,{\rm keV}$ (SASE)\\
        \hline
        \multirow{2}{*}{$\Delta\omega$} & $300\,{\rm meV}$ (self-seeded)\\
        & $10^{-3}\times\omega$ (SASE)\\
        \hline
        $\tau_{\rm FWHM}$ & $25\,{\rm fs}$ \\
        \hline
        \multicolumn{2}{| c |}{\bf ReLaX}\\
        \hline
        \hline
        $\lambda$ & $800\,{\rm nm}$\\
        \hline
        $W$ & $4.8\,{\rm J}$ \\
        \hline
        $\tau_{\rm FWHM}$  & $30\,{\rm fs}$\\
        \hline
        $w_{\rm FWHM}$  & $\#\times 1.3\,\upmu{\rm m}$\\
        \hline
    \end{tabular}
\end{adjustbox}
\caption{European XFEL and ReLaX parameters. The ReLaX focal width given here is for $f/\#$ focusing.}
\label{tab:params1}
\end{table}

\subsection{Conventional two-beam scenario}
\label{subsec:2beams}

\begin{figure}[h]
    \centering
    \includegraphics[width=0.95\textwidth]
    {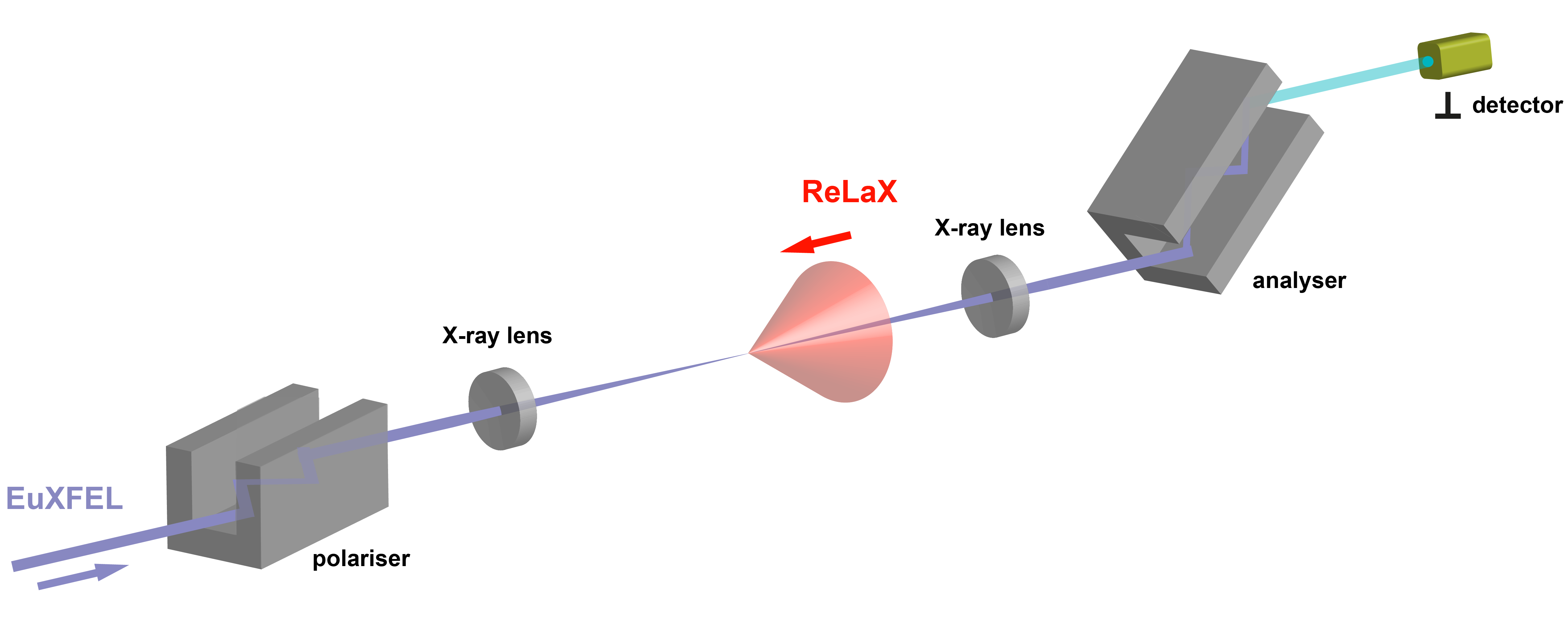}
    \vspace*{2mm}
    \caption{Schematic layout of the conventional scenario to measure vacuum birefringence. The XFEL beam is polarised with a channel-cut crystal, focussed down to the interaction point with the counter-propagating high-intensity laser, recollimated, and analysed with a second channel-cut crystal in crossed position, such that only the $\perp$-polarised component reaches the detector.}
    \label{fig:ConvBiref}
\end{figure}

The most extensively studied scenario in the present context is the conventional two-beam scenario \cite{Aleksandrov:1985}, which envisions the head-on collision of the XFEL beam acting as probe and a NIR high-intensity pump \cite{Heinzl:2006xc}.
It aims at measuring the birefringence phenomenon for probe light traversing a strong electromagnetic field that arises as a direct consequence of the effective nonlinear couplings of the electromagnetic field in \eqnref{eqn:HE1}: If the strong field introduces a preferred direction, such as, e.g., a unidirectional magnetic field or a linearly polarised laser field, the isotropy of the vacuum is broken. In turn, the vacuum polarised by the strong field effectively features two distinct transverse propagation eigenmodes for probe light associated with different refractive indices.
Correspondingly, if initially linearly polarised light (photon number $N$) having overlap with both of these modes is sent through such a field, a small fraction $N_\perp/N\ll1$ of its photons is scattered into a perpendicularly polarised mode, thereby effectively supplementing it with a small ellipticity.
The large photon number and the high polarisation degree of XFEL beams in conjunction with the available high-purity x-ray polarimetry, reaching polarisation purities down to the ${\cal P}=10^{-11}$ level for specific photon energies in the ${\cal O}(10)\,{\rm keV}$ regime \cite{Schulze:2022}, makes XFEL beams the ideal probe for such an experiment \cite{Schlenvoigt:2016jrd}.
At the same time, NIR  high-intensity lasers can reach the highest peak field strengths on macroscopic scales extending over spatial distances of ${\cal O}(1)\,\upmu{\rm m}$ and times of ${\cal O}(10)\,{\rm fs}$ which are well-compatible with the characteristic scales of XFEL pulses, making the best choice for the pump. In the collision of linearly polarised laser beams the number of polarisation-flipped signal photons $N_\perp$ is maximised for a relative polarisation angle of $\pi/4$.
As the attainable signal photon number scales with $(1-\cos\vartheta_{\rm coll})^4$, where $\vartheta_{\rm coll}$ is the collision angle of the two beams, the counter-propagating geometry is favoured; see Fig.~\ref{fig:ConvBiref} for a schematic illustration of the experimental setup.

The most advanced theoretical modelling of this specific scenario has used pulsed paraxial beams to describe the colliding laser fields. As the smallest XFEL beam waists that have been experimentally realised so far are of the order of $100\,{\rm nm}$ \cite{Mimura:2014}, the Rayleigh lengths that can be achieved for beams with a photon energy of $\omega_{\rm X}={\cal O}(10)\,{\rm keV}$ fulfil $z_{\rm R,X} \gtrsim {\cal O}(1)\,{\rm mm}$; cf. the definitions in the context of \eqnref{eqn:parax1} above.
This value is much larger than the Rayleigh range $z_{\rm R,L}={\cal O}(10)\,\upmu{\rm m}$ of a tightly focussed NIR laser beam. It is therefore an excellent approximation to formally send $z_{\rm R,X}\to\infty$ when determining the induced quantum vacuum signal, i.e., to adopt an {\it 
infinite Rayleigh length approximation} \cite{King:2010nka,King:2012aw,Gies:2017ygp,King:2018wtn,Karbstein:2021ldz} for the probe beam.
This significantly simplifies the calculation; cf. \cite{Karbstein:2021hwc} for the limitations of this approximation.
Modelling both laser fields as fundamental Gaussian beams at leading order in the paraxial approximation (multiplied by a Gaussian pulse envelope), the directional emission characteristics of the signal and the signal photon yield can be straightforwardly evaluated by standard computer algebra even for arbitrary collision angles and impact parameters \cite{King:2012aw,Dinu:2014tsa,Karbstein:2016lby}; for different measurement concepts see also \cite{Ataman:2018ucl,Formanek:2023mkx}.
Invoking further approximations, such as the inherently very small divergence of the signal in the x-ray domain and an {\it effective waist approximation} for the pump \cite{Karbstein:2018omb} (for the case of $\vartheta_{\rm coll}=\pi$), one obtains rather compact analytical scaling laws describing the dependence of the signal on the various parameters of the driving laser fields.
Using these scaling laws, one can quickly assess the effects of varying the characteristic parameters and thus identify promising parameter regimes prior to performing large-scale numerical simulations; cf. also the discussion below.

For a relative polarisation of $\pi/4$ between the colliding beams, the attainable number of polarisation-flipped signal photons per shot can then be approximately expressed as \cite{Karbstein:2018omb,Mosman:2021vua}
\begin{equation}
 N_{\perp} = N_{\rm X}\,\sqrt{\frac{3}{\pi}}\frac{(c_1-c_2)^2m^8}{\pi}
 \left(\frac{W_{\rm L}}{m}\frac{\omega_{\rm X}}{m}\right)^2\left(\frac{\lambdabar_e}{w_{0,{\rm L}}
 }\right)^4 \sqrt{g(0)g(\tfrac{w_{0,{\rm X}}}{w_{0,{\rm L}}})}
 \,\,\exp\left\{-2\left(\frac{r_0}{w_{0,{\rm X}}}\right)^2\left(1-\sqrt{\frac{g(\tfrac{w_{0,{\rm X}}}{w_{0,{\rm L}}})}{g(0)}}\right)\right\}\,, \label{eq:Nperp}
\end{equation}
where $c_1$ and $c_2$ are the low-energy constants introduced in \eqnref{eq:c_i}, $W_{\rm L}$ is the laser pulse energy of the pump,
$w_{0}$ denotes the beam waists,
$r_0$ is the transverse impact parameter, and we used the shorthand notation
\begin{align}
 &g(b):=\frac{1}{(1+2b^2)^2}\,F\left(\tfrac{4z_{{\rm R},{\rm L}}\sqrt{1+2b^2}}{\sqrt{\tau_{\rm X}^2+\tau_{\rm L}^2/2}},\tfrac{2(z_0-t_0)}{\sqrt{\tau_{\rm X}^2+\tau_{\rm L}^2/2}},\tfrac{\tau_{\rm X}}{\tau_{\rm L}}\right) ,\quad\text{with} \nonumber\\ &F(\chi,\chi_0,\rho):=\sqrt{\frac{1+2\rho^2}{3}}\,\chi^2\, {\rm e}^{2(\chi^2-\chi_0^2)} \int_{-\infty}^\infty {\rm d}K\,{\rm e}^{-K^2}\,\biggl|\sum_{s=\pm1} {\rm e}^{2s(\rho K-i\chi_0)\chi}\,
 {\rm erfc}\Bigl(s (\rho K
 -i\chi_0)+\chi\Bigr)\biggr|^2\,.
 \label{eq:F}
\end{align} 
The entire dependence of \eqnref{eq:Nperp} on the longitudinal beam parameters, namely the Rayleigh length of the pump $z_{{\rm R,L}}$, the $1/{\rm e}^2$ pulse durations $\tau_{\rm X}$, $\tau_{\rm L}$ and the spatiotemporal offset $z_0-t_0$ between the probe and pump foci in longitudinal direction, is encoded in the function $g(b)$ defined in \eqnref{eq:F}.
Note that in the limit of $w_{0,{\rm X}}\ll w_{0,{\rm L}}$ and $\{\tau_{\rm X},\tau_{\rm L}\}\gg z_{\rm R,L}$ we have $g(b)\sim(z_{\rm R,L}/\tau_{\rm L})^2$ \cite{Karbstein:2018omb}, and the form of \eqnref{eq:deltasquared} with characteristic distance $z\sim z_{\rm R,L}$ is recovered.
The dependence of the signal on the impact parameter $r_0$ is particularly relevant, as experiments have shown that focusing leads to an inevitable beam jitter characterised by spatial fluctuations of the order of the beam waist, i.e., $r_0 \sim w_{0,{\rm L}}$.
Another important signal parameter is the $1/{\rm e}^2$ radial divergence,
\begin{equation}
 \theta_{{\rm signal}}=\theta_{\rm X}\,\biggl(\frac{g(\tfrac{w_{0,{\rm X}}}{w_{0,{\rm L}}})}{g(0)}\biggr)^{-\frac{1}{4}}\,,
 \label{eq:divsig}
\end{equation}
where $\theta_{\rm X}=2/(\omega_{\rm X}w_{0,{\rm X}})$ is the radial divergence of the probe beam, and thus also the divergence of the background $N_{\perp,{\rm bgr}}\sim{\cal P}N_{\rm X}$ to be registered at the detector.

Equation~\eqref{eq:Nperp} predicts the maximum value of $N_\perp$ for optimal collisions ($r_0=z_0=t_0=0$) and fixed other parameters to be reached for small probe waists, $w_{0,{\rm X}} \ll w_{0,{\rm L}}$.
In this case, one obtains the polarisation-flip counts
\begin{equation}
    \begin{tabular}{c||c|c}
     $f$/\# & $f/1$ & $f/2$ \\
     \hline
  $N_\perp$   & $0.044$ & $0.0045$ 
\end{tabular} \; \; ,
\label{eq:Nperpmax}
\end{equation}
for an XFEL energy of $\omega_{\rm X}=9835\,{\rm eV}$ and the standard self-seeding parameters in Tab.~\ref{tab:params1} with $f/1$ ($f/2$) focusing.
Self-seeding is the option of choice for experiments aiming at the detection of polarisation-flip signals as the probe bandwidth $\Delta\omega$ is of the same order as the quite narrow acceptance bandwidth of crystal polarimeters. In contrast, the SASE bandwidth is typically much larger, cf. Tab.~\ref{tab:params1}.
In the determination of $N_\perp$ we also took into account that the quasi-channel-cut crystal acting as polariser increases the probe pulse duration from its original value given in Tab.~\ref{tab:params1} to $\tau_{\rm FWHM}=50\,{\rm fs}$ \cite{Karbstein:2021ldz}. 
Using ${\cal P}=10^{-11}$ \cite{Schulze:2022} and neglecting real-word imperfections (such as inevitable losses coming with the lenses used for focusing), the values in \eqnref{eq:Nperpmax} imply the associated signal-to-background ratios to be given by
\begin{equation}
    \begin{tabular}{c||c|c}
     $f$/\#  & $f/1$ & $f/2$ \\
     \hline
  $N_\perp/({\cal P}N_{\rm X})$   & $0.022$ & $0.0023$
\end{tabular} \; \; .
\label{eq:NperpbyPN}
\end{equation}
These values are less than unity so the signal is background-dominated. Moreover, in the limit considered, we have $\theta_{\rm signal} \approx \theta_{\rm X}$, and 
the dependence of the signal on the impact parameter $r_0$ simplifies to $N_\perp\sim\exp\{-2(r_0/w_{0,{\rm L}})^2\}$.

On the other hand, for matching waists, $w_{0,{\rm X}}= w_{0,{\rm L}}$, the signal is reduced by roughly a factor of $3$ to $N_\perp\simeq0.017$ ($0.0015$) for $f/1$ ($f/2$) focusing. At the same time the divergence of the signal, $\theta_{\rm signal} \approx 1.6\,\theta_{\rm X}$, becomes wider than that of the probe.
This can be intuitively understood as follows: For a counter-propagating geometry the transverse extent of the strong-field interaction region is effectively determined by the transverse profile of the product ${\cal E}_{\rm X}(x)\,{\cal E}^2_{\rm L}(x)$ of the focus field profiles of probe and pump, the Fourier transform of which governs the far-field distribution of the signal.
When $w_{0,{\rm X}} \gtrsim w_{0,{\rm L}}/\sqrt{2}$, the transverse extent of the signal emission region is smaller than the focal spot of the probe, and in turn $\theta_{\rm signal} > \theta_{\rm X}$; the factor of $\sqrt{2}$ is due to the linear (quadratic) dependency on the probe (pump) profile.
Furthermore, as $N_\perp \sim \exp\{-1.2(r_0/w_{0,{\rm L}})^2\}$ for $f/1$ focusing, the dependence on $r_0$ -- and thus the sensitivity to beam jitter in the experiment -- becomes milder.
Clearly, when $\theta_{\rm sig} > \theta_{\rm X}$, a measurement of the full signal requires collecting photons scattered outside the forward cone of the probe on the detector. However, as $\theta_{\rm X} \sim 1/w_{0,{\rm X}}$, the most tightly focussed probes make the highest demands on the diameter of the collection optics. 

The above simple comparison of just two different focusing options for the probe clearly illustrates that assessing the best choice of parameters for a feasible vacuum birefringence experiment is a nontrivial task. 
We emphasise that improved modelling accounting for the details of the laser fields actually available in experiments will become computationally demanding and challenging.

The possibility of achieving $\theta_{\rm sig} > \theta_{\rm X}$ suggests that angular cuts can be used to improve the signal-to-background ratio in experiment: Collecting only those photons scattered outside a minimum diffraction angle $\vartheta_{\rm min}$, the signal-to-background ratio can be enhanced to $N_\perp/({\cal P}N_{\rm X})|_{\vartheta>\vartheta_{\rm min}}>1$ at the expense of reduced absolute photon numbers. Following this route, it is even possible to achieve $N_\parallel/N_{\rm X}|_{\vartheta > \vartheta_{\rm min}} > 1$, such that one can hope to also measure the $\parallel$-polarised signal component \cite{Inada:2017lop,Karbstein:2019bhp}.
However, an experimental implementation of such a {\it diffraction-assisted measurement concept} \cite{King:2010kvw,King:2012aw} requires that any unwanted background, e.g., stray light from the lenses used for focusing, can be controlled and inhibited from reaching the detector, cf.\ \cite{Doyle:2021mdt} for a related background study at optical frequencies. This would necessitate setting up a dedicated filtering and imaging system along the lines outlined in Sec.~\ref{sec:experiment} for the dark-field scenario.

Finally, with regard to an actual experimental implementation we need to account for several non-ideal effects associated with the optical system: (i) In experiment each reflection at a diamond surface of the quasi-channel-cut polariser and analyzer comes with a reduction of the number of photons by about $2\,\%$ \cite{Karbstein:2021ldz} (the setup in Fig.~\ref{fig:ConvBiref} employs $8$ reflections in total).
(ii) The transmission of lenses preserving the polarisation purity can be estimated as $25\,\%$. (iii) The acceptance bandwidth of the polariser, $\Delta\omega_{\rm diamond}^{400} = 80\,{\rm meV}$, entails a reduction of its throughput by $\Delta\omega_{\rm diamond}^{400}/\Delta\omega = 80/300$ for self-seeding. In consequence, both the signal and background photon numbers, $N_\perp$ and ${\cal P}N_{\rm X}$, respectively, are reduced by a factor of $(0.98)^8 \times (0.25)^2 \times 80/300 \simeq 0.014$ from their ideal theoretical values stated in \eqnref{eq:Nperpmax} above.

While this keeps the signal-to-background ratios $N_\perp/({\cal P}N_{\rm X})$ given in \eqnref{eq:NperpbyPN} unaltered, it results in quite a drastic reduction of the signal photon numbers attainable per shot.
In particular, the best value attainable for $w_{0,{\rm X}} \ll w_{0,{\rm L}}$ then gives rise to just $N_\perp\simeq6.2\times10^{-4}$ ($6.4\times10^{-5}$) polarisation-flipped signal photons per shot reaching the detector for $f/1$ ($f/2$) focusing.
Assuming Poissonian statistics,
\begin{equation}
    n > \#^2\,\frac{1}{2}\Bigl[(N_{\rm sig}+N_{\rm bgr})\ln\Bigl(1+\frac{N_{\rm sig}}{N_{\rm bgr}}\Bigl)-N_{\rm sig}\Bigr]^{-1}=\#^2\,\frac{N_{\rm bgr}}{N_{\rm sig}^2}\Bigl[1+{\cal O}\Bigl(\frac{
    N_{\rm sig}
    }{N_{\rm bgr}}\Bigr)\Bigr]
    \label{eq:nshot}
\end{equation}
shots are required for a measurement with a significance of $\#\,\sigma$ \cite{Cowan:2012}, where $N_{\rm sig}$ and $N_{\rm bgr}$ are the number of signal and background events, respectively.
This implies that $n > \#^2\times7.5 \times 10^4$ ($\#^2 \times 6.9 \times 10^6$) optimal shots are required for a $\#\,\sigma$ confirmation of the vacuum birefringence signal with the presently available parameters at HED-HIBEF and $f/1$ ($f/2$) focusing. Given a repetition rate of $1 \,{\rm Hz}$, this translates into the requirement of more than $\#^2 \times 21 \,{\rm hours}$ ($\#^2 \times 80 \, {\rm days}$) optimal shots with zero spatiotemporal offset.

\subsection{Dark-field scenario}
\label{subsec:darkfield}

\begin{figure}[h]
    \centering
    \includegraphics[width=0.9\textwidth]{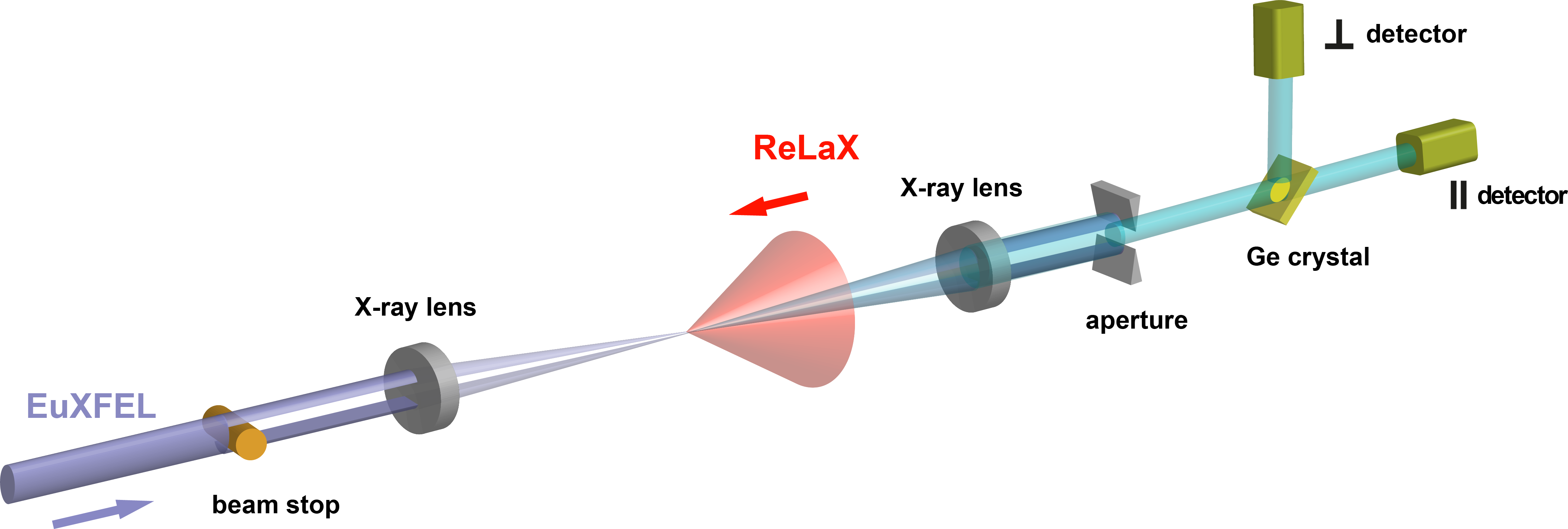}
    \vspace*{2mm}
    \caption{Schematic layout of the dark-field scenario.
    The XFEL is focussed with a beamstop creating a shadow in the converging (expanding) beam before (after) focus while retaining a central intensity peak in the focus where it collides with a counter-propagating high-intensity pump.
    X-ray optics image the beamstop to a matched aperture plane. 
    The effective interaction with the pump is strongly localized and limited to the vicinity of the probe focus. Hence, given that the overlap factor ${\cal E}_{\rm X}(x){\cal E}^2_{\rm L}(x)$ is sufficiently similar to that in the conventional scenario in Fig.~\ref{fig:ConvBiref}, a scattering signal is induced in the shadow. A crystal polariser directs its different polarisation components to separate detectors.}
    \label{fig:DarkField}
\end{figure}

A prospective variation of the conventional two-beam scenario detailed in Sec.~\ref{subsec:2beams} envisions modifying the probe beam such as to feature a shadow, or equivalently a {\it dark field} in the converging and expanding beam while retaining a central intensity peak in the focus where it is collided with the counter-propagating high-intensity pump \cite{Karbstein:2020gzg,Karbstein:2022uwf}.
In the experiment the shadow in the probe beam is generated by a well-defined beamstop inserted into the incident beam; see Fig.~\ref{fig:DarkField} for an illustration.
To demonstrate the underlying principle, we focus here on the case of a circularly symmetric shadow imprinted in the incident probe beam.
Experimentally, this annular beam approach was pioneered in \cite{Peatross:94,Zepf:1998} for the detection of weak nonlinear optics signals driven by high-intensity lasers.
By construction, the background is substantially reduced in the dark field, so that it may even be possible to access both polarisation components of the nonlinear vacuum response. To this end, the shadow in the probe beam is to be imaged onto a polarisation-sensitive detector employing an advanced filtering and imaging system.  This will consist of appropriately designed and placed beamstops and apertures such as to prevent any unwanted background from reaching the detector together with the signal.
We quantify the background scattered into the dark field by the {\it shadow quality} 
\begin{equation}
 {\cal S}=\frac{1}{N_{\rm X}}\int_{A_{\rm det}}{\rm d}^2x\,\frac{{\rm d}^2N_{\rm X}(x,y)}{{\rm d}x\,{\rm d}y}
 \label{es:shadowquality}
\end{equation}
that measures the fraction of the total number of input probe photons $N_{\rm X}$ registered by a detector of acceptance area $A_{\rm det}$ in the shadow in the absence of a quantum vacuum signal, i.e., for a vanishing pump field.
In a theoretically idealised calculation using ideal fat-top far-field profiles and neglecting diffraction at the obstacles and apertures put into the beam path, we have ${\cal S}=0$.
However, in any real experiment, we inevitably have ${\cal S}>0$, i.e., a finite background in the shadow, due to diffraction and scattering off the optical elements (cf. Sec.~\ref{subsec:DFexperiment} below).
The discussion of this set-up follows \cite{Karbstein:2022uwf} and more detail can be found in this work.

In the focus, the information about the far-field profile of the beam is encoded in a characteristic {\it Airy pattern} around the main peak.
The peak-field driven quantum vacuum nonlinearities producing the signal are sizable only in the vicinity of the overlapping beam foci. Thus, by ensuring that the transverse profile of the overlap factor ${\cal E}_{\rm X}(x)\,{\cal E}^2_{\rm L}(x)$ effectively matches the focus profile of a beam without shadow \cite{Karbstein:2023}, one can induce a signal component scattered into the shadow located in the expanding probe beam (cf. Sec.~\ref{subsec:2beams} above).

Note that this dependency immediately implies a signal in the dark field for $w_{0,{\rm X}} \gtrsim w_{0,{\rm L}}/\sqrt{2}$. In this case, the Gaussian pump intensity profile ensures that the Airy pattern of the probe in the focus is effectively damped out such that the transverse profile of the overlap factor ${\cal E}_{\rm X}(x)\,{\cal E}^2_{\rm L}(x)$ is characterised by a single Gaussian peak. This results in a Gaussian far-field distribution of the signal. At the same time, the fraction of the total induced signal that is scattered into the dark field decreases with $w_{0,{\rm L}}$, because the far-field divergence scales inversely with the source size.
On the other hand, it is clear that for $w_{0,{\rm X}} \ll w_{0,{\rm L}}$ the pump field appears as effectively constant on the transverse scales of variation of the probe. In this case, the dependence of the overlap factor ${\cal E}_{\rm X}(x)\,{\cal E}^2_{\rm L}(x)$ on the transverse coordinate $r$ is governed by the $r$ dependence of ${\cal E}_{\rm X}(x)$. Hence, in this limit, the dark field is also present in the far-field angular distribution of the signal. These considerations indicate that pump and probe waists of the same order maximise the signal scattered into the dark field.

In what follows, we discuss the dark-field scenario based on a few well-justified approximations. This allows us to provide simplified analytical expressions for the relevant quantities. (The full calculations accounting for the details of the probe beam can be found in \cite{Karbstein:2020gzg,Karbstein:2023}.)
To be specific, we limit ourselves to $w_{0,{\rm X}}\gtrsim w_{0,{\rm L}}/\sqrt{2}$ and make use of the fact that in this parameter regime, the focus profiles of both colliding laser fields can be well-approximated as Gaussian \cite{Karbstein:2023}.
In addition, we assume a relative polarisation of $\pi/4$ between the colliding beams and a detector of radial opening angle $\theta_{\rm det}$ placed centrally in the dark zone of the outgoing far field. The number $N_\perp^\bullet$ of $\perp$ polarised signal photons registered (per shot) by this detector is then well-approximated by \cite{Karbstein:2023}
\begin{equation}
 N_{\perp}^\bullet = N_\perp\, (1-\nu)^2\,\frac{1-{\rm e}^{-1}}{1+\nu}\left(1-{\rm e}^{-2\bigl(\theta_{\rm det}/\theta_{\rm signal}\bigr)^2}\right)\,,
 \label{eq:NperpDF}
\end{equation}
with $N_\perp$ and $\theta_{\rm signal}\leq\theta_{\rm in}$ as given in Eqs.~\eqref{eq:Nperp} and \eqref{eq:divsig}. For the same scenario, the complementary signal is somewhat larger, namely
\begin{equation}
    N_\parallel^\bullet=\biggl(\frac{c_1+c_2}{c_1-c_2}\biggr)^2\,N_\perp^\bullet\simeq13.4\,N_\perp^\bullet\,.
\end{equation} 
Here, $\nu=(\theta_{\rm in}/\theta_{\rm out})^2$ measures the fraction of the transverse area of the incident XFEL beam that is blocked by the beamstop. This blocking fraction can be parameterised in terms of the inner and outer radial divergences, $\theta_{\rm in}$ and $\theta_{\rm out}$, of the `hollow' probe beam featuring the central shadow \cite{Karbstein:2022uwf}, 
\begin{equation}
 \theta_{\rm out}=\frac{2}{\omega_{\rm X} w_{0,{\rm X}}}\sqrt{\frac{2(1-{\rm e}^{-1})}{1+\nu}} \quad\text{and}\quad \theta_{\rm in}=\sqrt{\nu}\,\theta_{\rm out}\,. 
 \label{eq:divs}
\end{equation}
Equation~\eqref{eq:NperpDF} follows upon integrating the differential number of signal photons derived in Refs.~\cite{Karbstein:2018omb,Mosman:2021vua} over solid angle. 
In addition, one has to account for the reduction of the peak field of the probe \cite{Karbstein:2020gzg} to match the central focus peak of a flat-top beam with a perfect central shadow in its far field.

From \eqnref{eq:NperpDF} we infer that the choice $\nu = \sqrt{5} - 2 \simeq 0.24$ maximises the signal in the shadow for optimal collisions.
We also choose $w_{0,{\rm X}} = w_{0,{\rm L}}/\sqrt{2}\simeq 0.7\,w_{0,{\rm L}}$ (which should be close to the optimal waist spot ratio \cite{Karbstein:2022uwf}) and an XFEL energy of $\omega_{\rm X}=8766\,{\rm eV}$ compatible with the use of a Germanium crystal to separate the polarisation components of the signal, cf. also Sec.~\ref{subsubsec:Ge} below for the details.
Adopting $\theta_{\rm det}=\theta_{\rm in}$ to collect the maximum number of signal photons scattered into the dark-field on the detector (for the parameters in Tab.~\ref{tab:params1} attainable with the self-seeding option), the second identity in \eqnref{eq:divs} then results in $\theta_{\rm in}\simeq28.3\,\upmu{\rm rad}/\#$, for $f/\#$ focusing. For these parameters, \eqnref{eq:NperpDF} predicts a signal yield which can be tabulated as
\begin{equation}
    \begin{tabular}{c||c|c}
     $f/\#$ & $f/1$ & $f/2$ \\
     \hline
  $N^\bullet_\perp$   & $0.0016$ & $1.2\times10^{-4}$
\end{tabular}\ .
\end{equation}
We also note that for a detector with half opening angle, i.e., for $\theta_{\rm det}=\theta_{\rm in}/2$, these photon numbers are reduced by a factor of $\approx 0.3$.
As in Sec.~\ref{subsec:2beams} we can also assess the dependence of the signal photon number on the impact parameter using \eqnref{eq:NperpDF}. For the present parameters and $f/1$ focusing we infer $N_\perp^\bullet\sim\exp\{-1.8(r_0/w_{0,{\rm L}})^2\}$.
Note, however, that the approximation invoked here accounts only for the central peak in the focus profile of the probe. It is therefore insensitive to the oscillatory behaviour of the signal induced by the Airy ring structure for $r_0\gtrsim w_{0,{\rm X}}$ and, in turn, is expected to somewhat overestimate the drop of the signal with $r_0$.

In the dark-field scenario, the polarisation purity of the probe photons does not need to be improved beyond the one supplied by the EuXFEL itself, which is of the order of $10^{-6}$ \cite{GELONI2015435}. Therefore, the analyser, which separates the differently polarised signal components scattered into the dark field, can work with a low number of reflections. This allows for a transmission bandwidth larger than that of the self-seeded XFEL and, at the same time, comes with a negligible influence on the probe pulse duration and a negligible loss at each reflection.
Using a \emph{single} reflection off Germanium, a polarisation purity of the order of $10^{-3}$ can be readily achieved. Employing \emph{several} such reflections, the polariser can be further enhanced, thus achieving a purity on par with the polarisation purity of the XFEL. 
In consequence, the losses associated with the setup in Fig.~\ref{fig:DarkField} are dominated by the lenses and not the analyser. Due to the relaxed polarisation purity requirement, standard beryllium lenses can now be used, each of which introduces a loss in the number of photons of about $50\,\%$. This reduces the signal and background photon numbers registered at the detector by a factor of $0.25$ and thus implies a reduction of the full signal scattered into the dark field to $N_\perp^\bullet \simeq 4.0 \times 10^{-4}$ ($3.0 \times 10^{-5}$) for $f/1$ ($f/2$) focusing. 
Assuming the Germanium polariser in Fig.~\ref{fig:DarkField} to achieve a polarisation purity of $10^{-6}$, the associated background in the dark field then consists of $N_{\rm bgr}={\cal S}N_{\rm X} \times 0.25 \times10^{-6}$ (${\cal S}N_{\rm X} \times 0.25$) XFEL photons for the $\perp$ ($\parallel$) polarised mode.
Identifying $N_{\rm sig}=N_\perp^\bullet$, and given that $N_{\rm sig} \ll N_{\rm bgr}$, \eqnref{eq:nshot} then implies the need of $n > \#^2 \times {\cal S} \times 3.2 \times 10^{11}$ ($\#^2 \times {\cal S} \times 5.5 \times 10^{16}$) optimal shots to achieve a $\#\,\sigma$ measurement of the $\perp$ polarised component of the nonlinear vacuum response with $f/1$ ($f/2$) focusing.
For $N_{\rm sig}=N^\bullet_\parallel$, the analogous values for the $\parallel$ polarised component are $n > \#^2 \times {\cal S} \times 1.8 \times 10^{15}$ ($\#^2 \times{\cal S} \times 3.0 \times10^{17}$).

To compare with the number of optimal shots needed for a measurement of the $\perp$ polarised signal in the conventional scenario of Sec.~\ref{subsec:2beams}, we refer back to the discussion below \eqnref{eq:nshot}. From this we conclude that one needs a shadow quality of ${\cal S}\lesssim 10^{-7}$ for the dark-field approach to match the precision of the conventional scenario.

Finally, we note that the measurement of both polarisation components of the nonlinear vacuum response allows to determine the low-energy constants $c_1$ and $c_2$ in \eqnref{eqn:HE1} individually, and not just their difference as in the conventional scenario.
Moreover, we emphasise that a simultaneous measurement of both components gives direct access to the ratio
\begin{equation}
    \frac{N_\perp}{N_\parallel}=\biggl(\frac{c_1-c_2}{c_1+c_2}\biggr)^2=\frac{9}{121}\biggl(1+\frac{260}{99}\frac{\alpha}{\pi}+{\cal O}\bigl(\alpha^2\bigr)\biggr)\,,
    \label{eq:flipratio}
\end{equation}
where we have also included higher-order (two-loop) contributions \cite{Ritus:1975cf,Dittrich:1985yb}.
Importantly, the ratio (\ref{eq:flipratio}) is independent of the parameters of the colliding beams and thus insensitive to fluctuations in experimental parameters such as spatiotemporal jitter or intensity fluctuations for sufficiently large photon numbers \cite{Karbstein:2022uwf}.

\subsection{Planar three-beam set-up}
\label{subsec:3beams}   

From a scattering point of view, vacuum birefringence is explained by polarisation flips without momentum transfer and hence corresponds to a LbL forward-scattering process. One can, however, relax this restrictive assumption and `open up' phase space, by considering genuine four-photon scattering processes with non-vanishing transfer of both energy and momentum. These are the fundamental processes that HED-HIBEF will measure. If two optical beams are employed instead of one, then each beam can couple one optical photon to the incoming XFEL photon, giving more control over the momentum and energy of the scattered signal photon 
\cite{King:2018wtn,Ahmadiniaz:2022nrv}. 
In particular, the scattered photon's momentum and energy can be shifted with respect to the XFEL probe photon so that the signal can be measured in a region of phase space with lower noise than the standard two-beam configuration. A disadvantage of this is that the number of scattered photons in this region is smaller than the total number of photons scattered in the two-beam configuration.

Let the field $F$ be given as a sum of the XFEL, $F_{\tsf{x}}$, two optical beams, $F_{1}$ and $F_{2}$, and the scattered photon $F_{\tsf{x}'}$, such that $F = F_{\tsf{x}} + F_{1} + F_{2} + F_{\tsf{x}'}$. Then, for the low-energy HE Lagrangian given in \eqnref{eqn:HE1}, the scattering matrix element, {$\tsf{S} = -i \int d^{4} x \, \mathcal{L}_{\trm{int}}$} contains many channels:
\bea 
\tsf{S} = \underbrace{\tsf{S}_{\tsf{x}11\tsf{x}'} + \tsf{S}_{\tsf{x}22\tsf{x}'}}_{\trm{x-ray}+\trm{one optical beam}} + \underbrace{\tsf{S}_{\tsf{x}12\tsf{x}'}}_{\trm{x-ray}+\trm{two optical beams}} + \cdots + \underbrace{\tsf{S}_{\tsf{x}\tsf{x}1\tsf{x}'}}_{\trm{x-ray merging}} + \cdots + \underbrace{\tsf{S}_{\tsf{x}1\tsf{x}'\tsf{x}'}}_{\trm{x-ray splitting}} + \cdots \; ,
\eea
where the subscripts on each of the amplitudes count the number of fields included (e.g. $\tsf{S}_{\tsf{x}11\tsf{x}'}$ involves one photon from the XFEL, two from optical beam $1$ and one scattered photon). If the optical beam is such that the field's tensor structure is constant as is the case for plane-wave-like fields (e.g. that of the paraxial Gaussian beam) then the kinematics of each amplitude can be illustrated by separating out the fields' spacetime dependency from their tensor structure, and writing kinematic factors $\Gamma(q)$ defined in terms of the momentum transfer $q$. For example, $\tsf{S}_{\tsf{x}11\tsf{x}'} = \mathfrak{S}\Gamma_{\tsf{x}11\tsf{x}'}(q)$ where $\mathfrak{S}$ is a geometrical factor depending on the polarisation of the photons and $\Gamma_{\tsf{x}11\tsf{x}'}(q) = \int F_{1}^{2}(x) \exp(-iq\cdot x) d^{4}x$. We note this has a form analogous to a far-field diffraction integral, where the x-ray photon `diffracts' on regions of the vacuum polarised by the intense optical beams \cite{DiPiazza:2006pr,King:2010nka}. These kinematic factors can be used to demonstrate the channels in the three-beam scenario; here it will suffice to consider monochromatic beams, e.g. $F_{j}(x) = F_{j} \cos (k_{j}\cdot x)$ for $j\in\{1,2\}$. Then we see that
\bea 
\Gamma_{\tsf{x}11\tsf{x}'}(q) \propto \underbrace{\delta\left(2k_{1}-q\right)}_{\trm{frequency upshift}}+\underbrace{\delta\left(2k_{1}+q\right)}_{\trm{frequency downshift}}+\underbrace{2\delta\left(q\right)}_{\trm{elastic}} \label{eqn:KF11}
\eea
i.e., the `x-ray+one optical beam' channel in principle includes frequency shifting parts. However, since the scattered photons with momentum $\ell'$ must obey the vacuum dispersion relation $\ell'\cdot\ell'=0$, the frequency shifting channels are effectively inaccessible \cite{King:2018wtn}. Compare this to the `x-ray+two optical beams' channel, for which the corresponding kinematic factor is
\bea 
\Gamma_{\tsf{x}12\tsf{x}'}(q) \propto \underbrace{\delta\left(k_{1}+k_{2}-q\right)}_{\trm{frequency upshift}}+\underbrace{\delta\left(k_{1}+k_{2}+q\right)}_{\trm{frequency downshift}}+\underbrace{\delta\left(k_{1}-k_{2}+q\right)}_{\trm{quasi-elastic}}+\underbrace{\delta\left(-k_{1}+k_{2}+q\right)}_{\trm{quasi-elastic}}. \label{eqn:KF12}
\eea
We see that by tuning $k_{1}$ and $k_{2}$ there is more control to scatter a photon with a momentum slightly different from the x-ray one, whilst simultaneously satisfying the on-shell condition $\ell'\cdot\ell'=0.$ It is indeed possible to pick three photons with not too different momenta and specific collision angles to access the frequency up- and down-shifting channels \cite{Bernard:2000ovj,Lundstrom:2005za,Lundin:2006wu,Gies:2017ezf,King:2018wtn,Aboushelbaya:2019ncg,Ahmadiniaz:2022nrv,Berezin:2024fxt}. However, for the x-ray + two optical photon collision at HED-HIBEF, only the `quasi-elastic' channels (of absorbing one photon from an optical beam and emitting one photon to an optical beam) are kinematically accessible. Furthermore if the absorbed and emitted photons have different momenta, the reaction is only accessible if one allows for the pulses having a finite bandwidth $\Delta$, i.e., for non-monochromatic beams, leading to a momentum transfer in the scattering process of $q = \pm(k_{1}-k_{2}) + \Delta$. Although labelled `quasi-elastic', these channels can support a measurable transverse momentum shift and, if one of the optical lasers is frequency doubled, also an energy shift in the scattered photons; see Fig.~\ref{fig:ralfFigs}.
This `three beam' kinematic picture can also be used to model frequency shifting in collisions of two beams with wide bandwidths, where the two photon momenta, $k_{1}$ and $k_{2}$, take different values supported by the bandwidth \cite{King:2012aw,Sundqvist:2023hvw}.

\begin{figure}[h!!]
    \centering
    \includegraphics[width=0.4\linewidth]{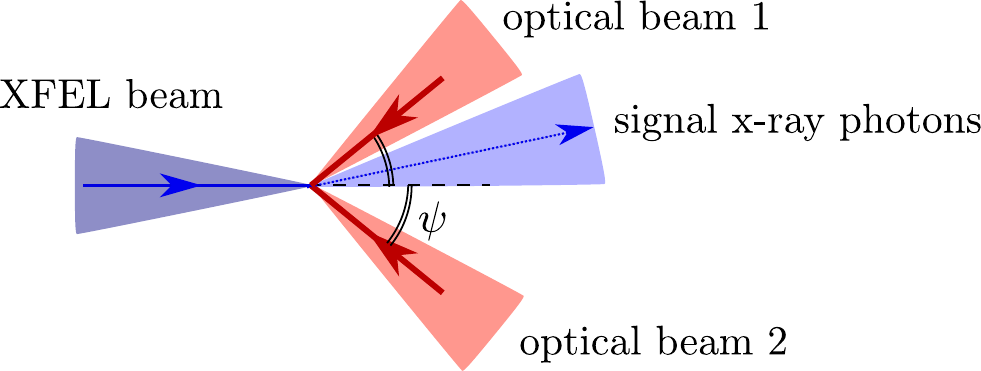}
    \caption{The planar three-beam configuration of a focussed XFEL beam colliding in a plane with two focussed optical beams at a collision angle $\psi$.}
    \label{fig:crossedBeamFig}
\end{figure}

\begin{figure}[h!!]
    \centering
    \includegraphics[width=0.6\linewidth]{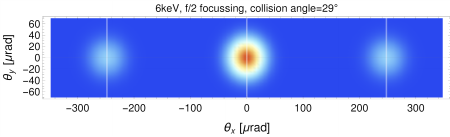}
    \caption{Example scattered photon signal for typical parameters at HED-HIBEF in the collision of the RELAX optical photons with the EuXFEL photons. The coordinates $\theta_{x,y}$ are the scattering angles in and perpendicular to the collision plane respectively.}
    \label{fig:crossedBeamExample1}
\end{figure}

\begin{figure}[h!!]
    \centering
    \includegraphics[width=0.8\linewidth]{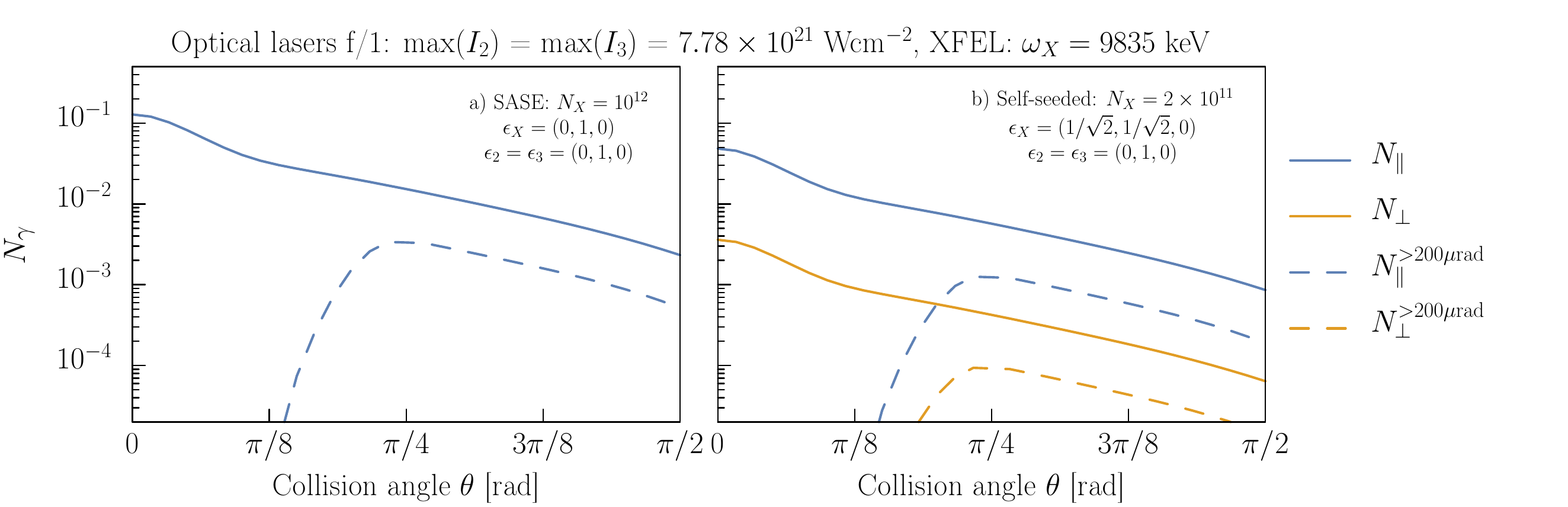}
    \caption{The number of scattered photons, parallel ($N_{\parallel}$) and perpendicular ($N_{\perp}$) to the XFEL probe, as a function of beam collision angle, plotted for different XFEL parameters. The probe propagates along the $z$ axis, and the collision is in the $x$-$z$ plane. The dashed curves ($N^{>200\upmu\trm{rad}}_{\parallel,\perp}$), refer to the signal falling on the detector outside a central exclusion region of radius $200\,\upmu\trm{rad}$. The SASE and self-seeded options are taken from \tabref{tab:params1}. Left: Example results for total scattered photons. Right: Photon scattering and birefringence (polarisation flip).}
\label{fig:crossedBeamExample2}
\end{figure}

Here we focus on a planar collision geometry for the three beams, depicted in \figref{fig:crossedBeamFig}.
Using the kinematic factors in \eqnref{eqn:KF11} and \eqnref{eqn:KF12} we can reinterpret the `x-ray+one optical beam' and `x-ray+two optical beams' scattering matrix channels in terms of positions in the detector plane,
\bea 
\tsf{S} \approx \underbrace{\tsf{S}_{\tsf{x}11\tsf{x}'} + \tsf{S}_{\tsf{x}22\tsf{x}'}}_{\trm{central peak}} + \underbrace{\tsf{S}_{\tsf{x}12\tsf{x}'}}_{\trm{side peaks}} \; . 
\label{eqn:Smin1}
\eea
Here we neglect all the channels that are nonlinear in the x-ray photons as they are accompanied by a suppression factor, the ratio of optical to x-ray field strengths. The centre of the side-peaks can be easily found from \eqnref{eqn:KF12}, at a scattering angle approximately twice the ratio of the transverse optical photon momentum to the x-ray momentum: $\theta_{x} = \pm \arctan(2\omega_{\tsf{optical}} \sin\psi/\omega_{\tsf{x-ray}})$, where $\theta_{x}$ is the scattering angle in the plane of the collision. The widths of the peaks can be ascertained by considering the bandwidth of the lasers. The number of scattered photons $\tsf{N}$ is calculated using $\tsf{N} = V\int \frac{d^{3}\ell'}{(2\pi)^{3}} |\tsf{S}|^{2}$, where $\ell'$ is the scattered photon momentum, and only the terms in \eqnref{eqn:Smin1} are included in $\tsf{S}$. The beams are modelled using the leading order paraxial Gaussian solution from \eqnref{eqn:parax1}, and parameters for the optical and EuXFEL beams taken from \tabref{tab:params1}. An example of the distribution of scattered photons is given in \figref{fig:crossedBeamExample1}, which assumes perfect alignment of the three beams, where the grid lines indicate the predicted central positions of the side peaks. 
\figref{fig:crossedBeamExample1} demonstrates the key utility of the three-beam set-up: the transverse momentum kick given to the signal photons can potentially be used to direct the signal into a region with low noise, albeit at the cost of a reduced signal. The larger the collision angle, the larger the momentum kick, but the lower the signal. This reduction is plotted in \figref{fig:crossedBeamExample2}. At small collision angles the three peaks constructively interfere since they are in the same region of phase space. As the collision angle is increased, the peaks separate and the signal reduces, while constructive interference becomes less pronounced.

To estimate the number of detectable photons, we exclude a central portion of the detector which will be dominated by the probe XFEL photons. From knowledge of the XFEL beam and numerical simulations (such as those detailed in \secref{sec:diffractive-simulations}) we determine that a $200\,\upmu\trm{m}$ radius for the central exclusion region on the detector would be sufficient. The entailed reduction in signal photon count is demonstrated in \figref{fig:crossedBeamExample2}. On the one hand, we would like a larger transverse momentum kick of the scattered photons so that the optical beam opening angle does not need to be made so large, and the detectable signal remains sufficient. This would suggest using lower energy XFEL photons, e.g. the $6\,\trm{keV}$ mode. On the other hand, the cross-section, $\sigma$, has a strong scaling with centre-of-mass energy, $\omega_{\ast}$ as $\sigma\propto \omega^{6}_{\ast}$, and so to maximise the total number of photons scattered, we would use the higher energy XFEL photons e.g. the $12.9\,\trm{keV}$ mode. From our analyses using the HED-HIBEF parameters, we find that the $9.8\,\trm{keV}$ mode is a good compromise in this planar three-beam scenario.

Some example results are given in Table~\ref{tab:threeBeamres}. Due to the bandwidth of the analyser being much narrower than the bandwidth of the XFEL beam in SASE operation, for observing birefringence, we use the self-seeded parameters for which the XFEL has a bandwidth narrow enough such that scattered photons are accepted by the analyser. 

\begin{center}
\begin{table}[h!!]
\begin{tabular}{|l||c|c|c|c|c||}
\hline
Crossed-beam set-up & Focussing & $\omega_{\tsf{x-ray}}$ & $N_{\parallel}$ & $N_{\perp}$ & $N_{\tsf{total}}$\\
\hline
\hline
`Momentum kick' & $f/2$ & 12.9\,\trm{keV} & - & - & $42 \times 10^{-3}$ \\
`Momentum kick' & $f/1$ & 12.9\,\trm{keV} & - & - & $180 \times 10^{-3}$ \\
\hline
`Momentum kick'+200$\upmu$m hole & $f/2$ & 12.9\,\trm{keV} & - & - & $1.5  \times 10^{-3}$ \\
`Momentum kick'+200$\upmu$m hole & $f/1$ & 12.9\,\trm{keV} & - & - & $3.7 \times 10^{-3}$ \\
\hline
\hline
Crossed beam + biref. & $f/1$ & $9.8\,\trm{keV}$ & $49\times 10^{-3}$ & $3.5 \times 10^{-3}$ & - \\
\hline
Crossed beam + biref.+200$\upmu$m hole & $f/1$ & $9.8\,\trm{keV}$ & $1.3\times 10^{-3}$ & $0.1\times 10^{-3}$ & - \\
\hline
\end{tabular}
\caption{Example number of signal photons scattered in collision of XFEL probe with two optical beams.}\label{tab:threeBeamres}
\end{table}
\end{center}

The planar three-beam scenario is a set-up of interest for BIREF@HIBEF. As the dark-field scenario has been selected for priority access beam time at the EuXFEL, we omit here a full discussion of the effect of optical components on the measurability of the numbers of photons detailed in Table~\ref{tab:threeBeamres}, and how the HE low-energy constants $c_{1}$ and $c_{2}$ can be determined in this set-up.

\subsection{Other opportunities: Coulomb-assisted birefringent scattering}
\label{subsec:other}

In the preceding sections, we have rigorously examined a variety of scenarios for detecting vacuum birefringence, each presenting unique challenges and insights. These explorations spanned across diverse experimental setups, employing a range of pump and probe fields to elucidate the nature of vacuum birefringence. These varying approaches, each with its own merits and limitations, collectively contribute to a better understanding of the phenomenon. These options are concisely summarised in Table~\ref{tab:pump-probe}. In what follows, we first provide an overview of the available choices for both pump and probe fields. Then, we venture into a new direction by combining two options for the pump field to explore a birefringent signal. 

\begin{table}[ht]
\centering
\begin{tabular}{|l|c|r|} 
\hline
\multicolumn{3}{|c|}{Pump Field}\\
\hline
\hline
Magnetic Field & Laser Focus & Nuclear Coulomb Field \\ 
\hline
$\mathcal{O}(10^{-9}B_{\mathrm{S}})$ & $\mathcal{O}(10^{-4}E_{\mathrm{S}})$ & $\mathcal{O}(E_{\mathrm{S}})$ \\ 
\hline
$\delta n=\mathcal{O}(10^{-22})$ & $\delta n=\mathcal{O}(10^{-11})$ & $\delta n=\mathcal{O}(10^{-2})$ \\
\hline
\multicolumn{3}{|c|}{Field Strength $\rightarrow$ \hspace{\fill} $\leftarrow$ Interaction Volume} \\
\hline  
\noalign{\vskip 3mm}
\hline
\multicolumn{3}{|c|}{Probe Field}\\
\hline
\hline
Optical Laser & XFEL & $\gamma$-Ray \\
\hline
$\mathcal{O}(\mathrm{eV})$ & $\mathcal{O}(\mathrm{keV})$ & $\mathcal{O}(\mathrm{MeV})$ \\
\hline
$N=\mathcal{O}(10^{20})$ & $N=\mathcal{O}(10^{11})$ & $N=\mathcal{O}(1)$ \\
\hline
\multicolumn{3}{|c|}{Wave number $\rightarrow$ \hspace{\fill} $\leftarrow$ Photon Number} \\
\hline
PVLAS, BMV, \dots & HED-HIBEF & Delbr\"uck \\
\hline
\end{tabular}

\caption{Summary of choices for pump and probe fields in vacuum birefringence detection. Increasing the strength of the pump fields results in a smaller interaction volume, while a higher wave number in the probe fields leads to fewer photons per shot.}
\label{tab:pump-probe}
\end{table}

Table~\ref{tab:pump-probe} presents three distinct choices each for pump and probe fields. In the case of the pump field, one can opt for a static magnetic field, laser focus, or the electric field of a nucleus, each with its own set of pros and cons. For example, the first column details the characteristics of a static magnetic field, where we achieve a field strength of several Tesla. This leads to an extremely small change in the vacuum refractive index ($\delta n$), and the other columns provide similar metrics for alternative pump field options. Regarding the probe field, there are also multiple choices. Take the optical laser, for instance: it generates a large number of photons $N$ but with relatively low energy $\mathcal{O}({\rm eV})$. The table in Table~\ref{tab:pump-probe} illustrates various experiments that have utilised different combinations of pump and probe fields. For instance, as detailed in \secref{sec:intro}, experiments like PVLAS \cite{Ejlli:2020yhk} and BMV \cite{Battesti:2007} employed the options detailed in the first column: a static field as the pump and an optical laser as the probe. HED-HIBEF is designed to follow the approach outlined in the second column, using a focussed array of several optical lasers for the pump field and x-rays for the probe. Additionally, for the measurement of Delbr\"uck scattering \cite{Jarlskog:1973aui,Schumacher:1975kv} have utilised the electric field of a nucleus as pump field and high-energy gamma photons as the probe; for more details see \cite{Ahmadiniaz:2020lbg}.
\begin{figure}[ht]
    \centering
    \includegraphics[width=0.5\linewidth]{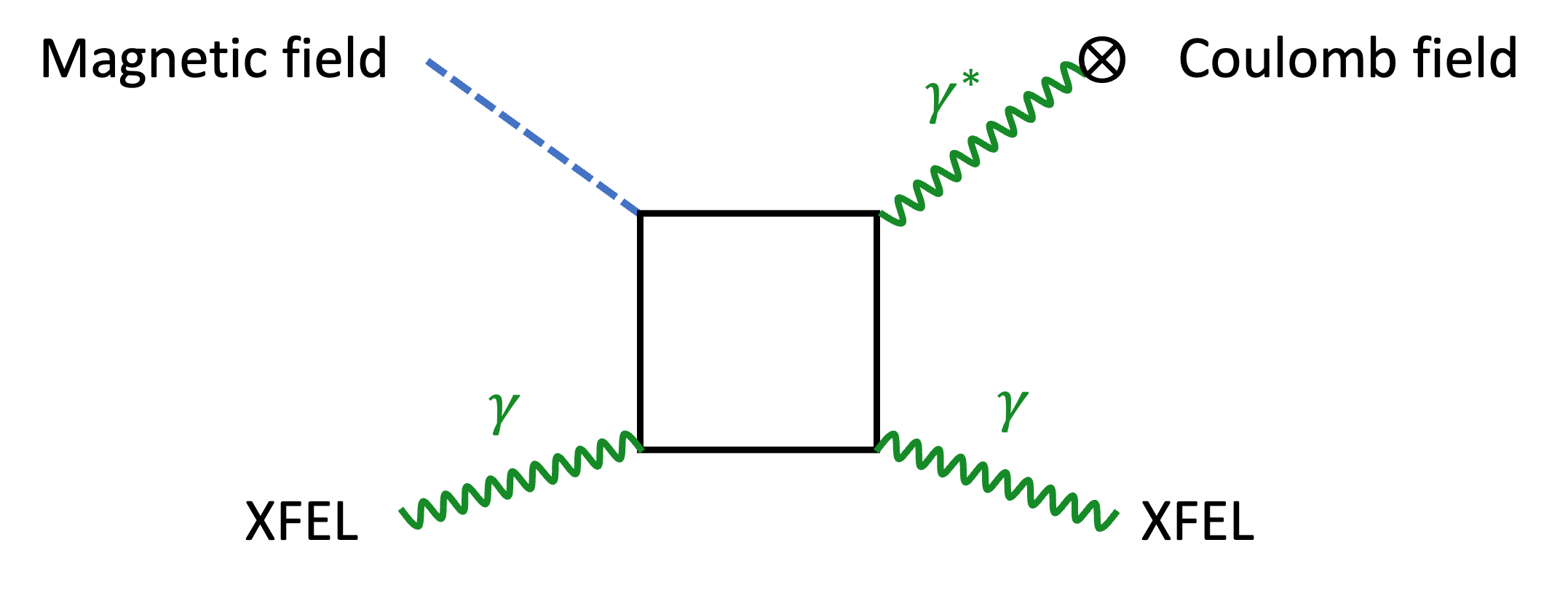}
    \caption{Feynman diagram for Coulomb-assisted birefringence}
    \label{fig:C_assisted}
\end{figure}
In this section, we utilise the combination of a nuclear Coulomb field, denoted as ${\bf E}^{\rm ext}$, and the strong magnetic field ${\bf B}^{\rm ext}$ provided by a focussed high-intensity laser pulse as pump field. This is coupled with an XFEL beam, which serves as the probe field, see  \cite{Ahmadiniaz:2020kpl,Ahmadiniaz:2020lbg}. An illustrative Feynman diagram for this process resembles the one shown in Fig. \ref{fig:DELBRUECK} (a), with the key distinction being that one of the Coulomb fields is replaced by an interaction with an external magnetic field ${\bf B}^{\rm ext}$, see Fig \ref{fig:C_assisted}. This scenario may offer several advantages: the nuclear Coulomb field adds a significant momentum transfer $\Delta{\bf k}$ to the flip of the XFEL polarisation \cite{Ahmadiniaz:2020kpl,Ahmadiniaz:2020lbg}. Compared to the past experiments measuring Delbr\"uck scattering at photon energies $O(100\,\trm{MeV})$ \cite{Jarlskog:1973aui,Schumacher:1975kv,Akhmadaliev:1998zz}, we obtain a large interaction volume whose space-time scale is set by the momentum transfer and thus exceeds the Compton wavelength. Since all involved field strengths are sub-critical, hence well below the Schwinger limit, we can use the low-field expansion of the HE Lagrangian in \eqref{eqn:HE1} with the low-energy constants $c_1$ and $c_2$ given in \eqref{eq:c_i}.
In the next step, we decompose the electric and magnetic fields in Eqs.~\eqref{eqn:HE1} and \eqref{invs} as ${\bf E}\to{\bf E}+{\bf E}^{\rm ext}$ and ${\bf B}\to{\bf B}+{\bf B}^{\rm ext}$ and assume that the magnetic field ${\bf B}^{\rm ext}$ is approximately constant. In addition to the static Coulomb field ${\bf E}^{\rm ext}$ of the nucleus, we have space-time dependent XFEL fields ${\bf E}$ and ${\bf B}$. After inserting the field decomposition into \eqnref{eqn:HE1} we obtain the effective Lagrangian for the XFEL field
\begin{eqnarray}
\mathcal{L}_{\rm XFEL} = \frac{1}{2}\Big[{\bf E}\cdot\epsilon^{\rm ext}\cdot{\bf E}+{\bf B}\cdot(\mu^{-1})^{\rm ext}\cdot{\bf B}+{\bf E}\cdot\Psi^{\rm ext}\cdot{\bf B}\Big]\, ,
\label{xfel-eff}
\end{eqnarray}
where we have introduced the notation ${\bf E}\cdot\epsilon^{\rm ext} \cdot{\bf E} \equiv E_i \epsilon_{ij}^{\rm ext} E_j$ etc.\ for the quadratic forms in the Lagrangian $\eqref{xfel-eff}$. Its Hessian defines the dynamical vacuum response functions \cite{Baier:1967zzc}, i.e.\ the permittivity tensor, $\epsilon_{ij}^{\rm ext}$, the permeability tensor, $(\mu^{-1}_{ij})^{\rm ext}$, and the magneto-electric coupling tensor, $\Psi_{ij}^{\rm ext}$, all of which are functions of the pump fields, ${\bf E}_{\rm ext}$ and ${\bf B}_{\rm ext}$,
\begin{eqnarray}
  \epsilon_{ij}^{\rm ext}&=&\frac{2\alpha^2}{45 m^4}\Big[8E^{\rm ext}_i E^{\rm ext}_j+14 B^{\rm ext}_iB^{\rm ext}_j+\delta_{ij}(({\bf E}^{\rm ext})^2-({\bf B}^{\rm ext})^2)\Big]\, ,\nonumber\\
  (\mu_{ij}^{-1})^{\rm ext}&=& \frac{2\alpha^2}{45 m^4}\Big[8B^{\rm ext}_i B^{\rm ext}_j+14 E^{\rm ext}_i E^{\rm ext}_j +\delta_{ij}(({\bf B}^{\rm ext})^2-({\bf E}^{\rm ext})^2) \Big]
  \, ,\nonumber\\  
  \Psi_{ij}^{\rm ext}&=&\frac{2\alpha^2}{45 m^4}\Big[-8E_{i}^{\rm ext}B_j^{\rm ext}+14 B^{\rm ext}_i E^{\rm ext}_j+14\delta_{ij}({\bf E}^{\rm ext}\cdot{\bf B}^{\rm ext})\Big]\, .
\end{eqnarray}
These tensors characterise the vacuum polarisability, which varies with the type of pump laser used as detailed in Table~\ref{tab:pump-probe}. The equations of motion derived from \eqnref{xfel-eff} turn into macroscopic Maxwell equations in medium upon introducing the electric displacement field ${\bf D} =  \epsilon ^{\rm ext}\cdot {\bf E} +\Psi^{\rm ext} \cdot {\bf B}$ and the magnetic displacement field ${\bf H} =  -( (\mu^{-1})^{\rm ext} \cdot {\bf B} +(\Psi^{\rm ext})^T \cdot {\bf E})$. In the usual manner, a suitable combination yields the inhomogeneous wave equation
\begin{eqnarray}
    \Box{\bf D}=\nabla\times[\nabla\times{\bf D}]+\partial_t[\nabla\times{\bf H}]={\bf J}^{\rm eff}\, .
\end{eqnarray}
The resulting wave equation can be analysed using standard Green's function methods (see e.g. \cite{Marklund:2006my}).
The source term ${\bf J}^{\rm eff}$, owing to its relatively small magnitude, permits the use of the Born approximation to solve the equation. To proceed we decompose the XFEL electric displacement field ${\bf D}$ into an incoming plane wave ${\bf D}^{\rm in}$ and a small scattering component ${\bf D}^{\rm out}$. Assuming a stationary time-dependence of ${\rm e}^{-i\omega t}$ for the XFEL, we arrive at a Helmholtz equation \cite{Ahmadiniaz:2020kpl}.
This equation can be solved using Green function methods. Focusing on the long-distance behaviour, we find the following scattering amplitude \cite{Jackson:1998nia},
\begin{equation}
    \mathfrak{U}=\frac{1}{4\pi\mid\bf{D}^{\rm in}_\omega\mid}\,{\bf e}_{\rm out}\cdot\int d^3r\,\exp\{-i\,{\bf k}_{\rm out}\cdot {\bf r}\}\,{\bf J}^{\rm eff}_\omega
\end{equation}
where ${\bf k}_{\rm out}$ is the wave-number and ${\bf e}_{\rm out}$ the polarisation unit vector of the scattered XFEL radiation. The scattering amplitude straightforwardly yields the differential cross section, $d\sigma/d\Omega=\mid\mathfrak{U}\mid^2$. For Coulomb assisted Delbr\"uck scattering, cf.\ Fig.~\ref{fig:C_assisted}, the amplitude can be expressed as
\begin{eqnarray}
    \mathfrak{U}&=&\frac{\omega^2}{4\pi}\int d^3r \, 
    {\rm e}^{i\Delta{\bf k} \cdot {\bf r}} \big[{\bf e}_{\rm out}\cdot\epsilon^{\rm ext}\cdot{\bf e}_{\rm in}+{\bf e}_{\rm out}\cdot\Psi^{\rm ext}\cdot({\bf n}_{\rm in}\times{\bf e}_{\rm in})+{\bf e}_{\rm in}\cdot\Psi^{\rm ext}\cdot({\bf n}_{\rm out}\times{\bf e}_{\rm out})
    \nonumber\\
    &&\hspace{8cm}\!
    +\,({\bf n}_{\rm out}\times{\bf e}_{\rm out})\cdot(\mu^{-1})^{\rm ext}\cdot({\bf n}_{\rm in}\times{\bf e}_{\rm in})\big]\, .
    \label{u-scattering}
\end{eqnarray}
Here ${\bf n}_{\rm in}={\bf k}_{\rm in}/\omega$ and ${\bf n}_{\rm out}={\bf k}_{\rm out}/\omega$ are the initial and final propagation directions, respectively, while ${\bf e}_{\rm in}$ and ${\bf e}_{\rm out}$ are the corresponding polarisation vectors; $\Delta{\bf k}={\bf k}_{\rm in}-{\bf k}_{\rm out}$ is the momentum transfer or recoil. \figref{fig:C_assisted} suggests a non-vanishing contribution of the magneto-electric tensor $\Psi^{\rm ext}$, i.e. the second and the third terms of \eqnref{u-scattering}. Since the external magnetic field ${\bf B}^{\rm ext}$ provided by an intense laser focus is approximately constant, the spatial integral in \eqref{u-scattering} yields the Fourier transform of the nuclear Coulomb field, ${\bf E}^{\rm ext}={\bf e}_r Ze/(4\pi r^2)$ ($Z$ being the atomic number), i.e., $\int d^3r\,{\rm e}^{i\Delta{\bf k}\cdot{\bf r}}{\bf e}_rZe/(4\pi r^2)=i Ze\Delta{\bf k}/|\Delta{\bf k}|^2$.
    
Note that the spatial integration is effectively cut off by the momentum transfer $\Delta{\bf k}$, implying a large interaction volume extending over many XFEL wavelengths for small scattering angles (near-forward direction), $\mid{\bf n}_{\rm in}-{\bf n}_{\rm out}\mid\ll1$. However, when the effective interaction volume becomes of the order of the spatiotemporal scales of variation of the optical laser pulse, the approximation of a constant magnetic field can now longer be justified; see also \cite{Lee:2017nug}. 

Let us consider forward scattering and focus on the leading order contribution $\sim 1/|\Delta{\bf k}|$.  We thus approximate ${\bf n}_{\rm in}\approx{\bf n}_{\rm out} \equiv {\bf n}$ and consider the birefringent signal where ${\bf e}_{\rm in}$ and ${\bf n}$ are (nearly) orthogonal, i.e., ${\bf e}_{\rm in}\approx\mp{\bf n}\times {\bf e}_{\rm out}$ and  ${\bf e}_{\rm out}\approx\pm{\bf n}\times {\bf e}_{\rm in}$. (The $\mp$ and $\pm$ symbols refer to two possible orientations of ${\bf e}_{\rm in}$ and ${\bf e}_{\rm out}$: either parallel or anti-parallel to the cross product of ${\bf n}$ with the other polarisation vector.) This simplifies the integrand in \eqnref{u-scattering}, and the birefringent amplitude is finally given by 
\begin{equation}
    \mathfrak{U}_{\perp}^{\Psi^{\rm ext}}=\pm i \, \frac{12\alpha^2}{45m^4} \, \frac{Ze}{(\Delta{\bf k})^2}\frac{\omega^2}{4\pi}
    \left[({\bf e}_{\rm out}\cdot{\bf B}^{\rm ext})({\bf e}_{\rm out}\cdot\Delta{\bf k})-({\bf e}_{\rm in}\cdot{\bf B}^{\rm ext})({\bf e}_{\rm in}\cdot\Delta{\bf k})\right] \, .
\end{equation}
We maximise the birefringence signal by aligning $\Delta {\bf k}$ with ${\bf B}^{\rm ext}$ and ${\bf e}_{\rm in}$ (or ${\bf e}_{\rm out}$).
Assuming a strong magnetic field of order $B^{\rm ext}=10^6 \;{\rm T}$ (which can be achieved through laser focusing) and a large XFEL frequency of $\omega=24\;{\rm keV}$ (which involves additional experimental challenges), the birefringence signal is encoded in the differential cross section in forward direction,
\begin{equation}
    \frac{d\sigma_\perp^{\Psi^{\rm ext}}}{d\Omega}=\mid\mathfrak{U}_\perp^{\Psi^{\rm ext}}\mid^2 \approx 10^{-25}\frac{Z^2}{(\Delta{\bf k})^2}\, .
    \label{cross-for}
\end{equation}
The magnitude of the signal \eqnref{cross-for} is comparable to the other proposals for the vacuum birefringence experiments discussed above. In the Coulomb-assisted scenario, there are two main enhancement factors: first, the large number, $\mathcal{O}(10^{12})$, of polarised XFEL photons; second, the large number $N$ of nuclei. In one possible scenario, a cubic carbon cluster with an edge length of $100\,{\rm nm}$ is ionised completely using the pre-pulse of a high-intensity laser of order $\mathcal{O}(10^{22}\,{\rm W/cm^2})$. For small momentum transfer (of order $\rm eV$), PIC simulations show that the amplitudes of $N=10^8$ nuclei would have the same phase and add up coherently, thus acting as one giant nucleus of charge $Z_{\rm eff}=NZ$. The birefringent signal would be of the order of $\mathcal{O}(10^{-5})$ photons per shot and thus could be detected in complete analogy
with the scenarios discussed above, see also \cite{Battesti:2018bgc,Heinzl:2006xc,Inada:2017lop,Yamaji:2016,Schlenvoigt:2016jrd}.  For a more detailed discussion, we refer to the recent publications \cite{Ahmadiniaz:2020kpl,Ahmadiniaz:2022mcy}, which also address a number of potential background processes, including nuclear Thomson scattering, nuclear resonances, Delbr\"uck scattering, and electronic Compton scattering. Among these processes, electronic Compton scattering seems the most important.

Another background atomic effect that could flip the polarisation of x-ray photons is birefringence and dichroism of atomic x-ray susceptibility due to the alignment of electrons in ions generated by strong-field ionisation by the polarised optical laser pulse. The x-ray dichroism of ionised krypton gas was experimentally demonstrated in \cite{young_x-ray_2006}, \cite{southworth_k_2007}, \cite{goulielmakis_real-time_2010} and theoretically studied in \cite{santra_spin-orbit_2006}, \cite{rohringer_multichannel_2009} for x-rays resonant to transition between the 1s core atomic orbital and 4p$_{3/2}$ valence atomic orbital where an aligned electron hole was produced.  From these studies, the probability of polarisation flipping can be estimated as $\sim [\sin(2\theta) n \lambda^2 L \Delta\rho_{\text{aniso}}]^2 A^2/(\Delta\omega^2 + \gamma^2/4)$, here $\theta$ is an angle between the polarisations of the optical and x-ray pulses; $n$ and $L$ are the concentration and the length of the gas sample; $\lambda$ is the x-ray wavelength, $\Delta\rho_{\text{aniso}}$ is the difference between probabilities of finding a valence orbital hole with distinct projection quantum numbers – this quantity describes the alignment of the valence hole; $A$, $\Delta \omega$ and $\gamma$ are the spontaneous emission rate, the detuning and the linewidth for the corresponding valence-to-core transition. For the case of strong-field ionised ${\rm Kr}^+{\rm 4p}^{-1}$, the alignment of the valence hole $\Delta\rho_{\text{aniso}}$ can reach the value of $64\%$, and for an atomic concentration 10$^{18}\,{\rm cm}^{-3}$, an interaction length of a few mm and an x-ray photon energy resonant to valence-to-core transition, the polarisation flipping probability could reach $\mathcal{O}$(10$^{-5}$). With decreasing gas concentration and increasing detuning, this probability rapidly drops off – thus avoiding the valence-to-core transition is needed to suppress the atomic birefringence background. From the AMO and plasma physics point of view, the high-purity x-ray polarimetry measurements described in this letter could provide unique insights into the anisotropy properties of atomic transitions.

\section{Plans for experimental implementation}
\label{sec:experiment}
Based on the expected signal and noise estimates presented in this paper, as well as on the experimental feasibility, the dark-field approach was selected to be pursued via the HIBEF user consortium priority access at the European XFEL. The first x-ray-only beamtime was allocated for March 2024.  It was devoted to carrying out a proof-of-principle experiment of the dark-field concept at XFEL. The outcomes of this campaign are currently being analysed and will determine the setup to be implemented for the actual discovery experiment.

\subsection{Conventional scenario}\label{subsec:Cexperiment}
The decisive bottle necks in the conventional scenario are the quality of the x-ray polarizers and the x-ray lenses. For the former, early works suggested that an extinction ratio on the order of $10^{-10}$ would be necessary to prove vacuum birefringence induced by a Petawatt-class laser. The extinction ratio is also known as polarisation purity. More detailed analyses, together with the fact that only $300\,{\rm TW}$ are available at the European XFEL, led to the insight that an extinction ratio of $10^{-12}$ is actually required for the x-ray polarimeter. Of equal importance is that the x-ray lenses do not compromise the polarisation purity. 

At the University of Jena and the Helmholtz Institute Jena, a research and development program to eliminate these two bottle necks has been started about 15 years ago. The polarizers are perfect crystals into which a trench is cut -- which is why they are referred to as channel-cut crystals or simply channel-cuts (CC). The wavelength is chosen such that the x-rays are Bragg-reflected at the (inner) walls of the channel at a Bragg angle of $45^\circ$, the Brewster angle for x-rays. At the Brewster angle, the polarisation component parallel to the diffraction plane ($p$-component) of the x-rays’ field is suppressed. Zig-zagging multiple successive reflections increases the suppression of the $p$-component. For some crystals attractive for x-ray polarimetry -- namely diamond -- cutting a channel is not an option. In this case, one would precision-mount two separate crystals perfectly in parallel, an arrangement known as artificial or quasi channel-cut (QCC).

As early as 2010, an extinction ratio of $1.5\times 10^{-9}$ at $6.457\,{\rm keV}$ photon energy and of $9.0\times 10^{-9}$ at $12.914\,{\rm keV}$ was demonstrated. The (400) and (800) Bragg reflexes of Si were used \cite{Marx:OC11}. Advances in the production and processing of the crystals and improvements in the brilliance of x-ray sources led relatively quickly to a further improvement in polarisation purity to $2.4\times 10^{-10}$ and $5.7\times 10^{-10}$ for the same x-ray wavelengths as above \cite{PhysRevLett.110.254801}. The originally communicated requirements on x-ray polarimetry to detect vacuum birefringence were thus largely fulfilled. While these requirements were to become more and more demanding, some of the opportunities arising from the advent of polarizers providing polarisation purities several (!) orders of magnitude better than the previous state-of-the-art were quickly exploited (e.g. \cite{PhysRevLett.110.254801,Schulze:JSR21,Schmitt:Optica21,Marx:NJP22}). Prominent applications were the detection of quantum optics and QED effects in the x-ray range \cite{Heeg:PRL13,Heeg:PRL15,Haber:NPhot16}.  

In order to achieve further improvements in polarisation purity, the limiting factors had to be identified \cite{Schulze:APL18,marx2015influence} and eliminated. These are the so-called Umweganregungen and the finite divergence of the x-rays. Strictly speaking, the latter is not a real problem, as it only occurs at the synchrotron and laboratory sources used for polarimeter development, whereas the divergence of the XFEL beam is negligible. With regard to the Umweganregungen, it turned out that these cannot be avoided in principle. However, there are certain highly symmetrical reflexes for which all of them interfere destructively. What is required, though, is a highly precise adjustment not only of the Bragg angle, but also of the azimuthal orientation of the polarimeter crystals, which is the degree of freedom that can be exploited by a rotation around the normal of the diffracting lattice planes. Novel methods have been developed for this task that enable quick adjustment of the polarimeter crystals and thus efficient use of the precious XFEL beam time. 

In parallel, another branch of the research and development of x-ray polarimeters was the use of diamond crystals. These are advantageous for several reasons. Notable among these are the small lattice constant, which allows the use of shorter wavelengths, the low atomic number, which is advantageous with regard to Umweganregungen, the high reflectivity of almost 100\% and, finally, the excellent thermal conductivity. First experiments were performed with relatively cheap CVD crystals in a QCC setup with only two reflections \cite{Bernhardt:APL16}. Nevertheless, a polarisation purity of $8.9\times 10^{-10}$ at $9.837\,{\rm keV}$ ((400) Bragg reflection) was obtained, limited by the divergence of the synchrotron radiation. With four crystals for each, the polarizer and the analyser, and some control on the beam divergence, another record polarisation purity of $1.4\times 10^{-10}$ was achieved \cite{Bernhardt:PRR20}.

\begin{figure}[h!!]
    \centering
    \includegraphics[width=0.75\linewidth]{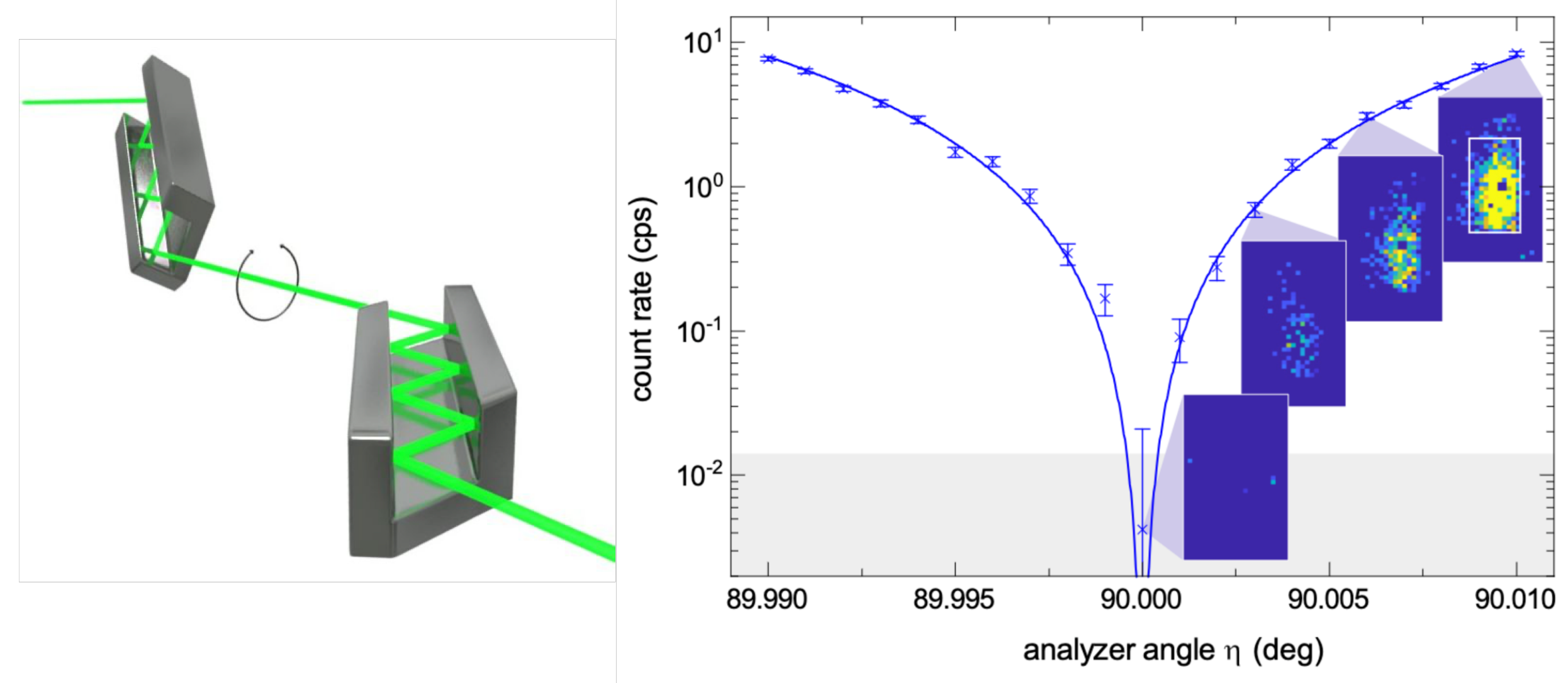}
    \caption{High-precision x-ray polarimetry. Left: The polarimeter consists of two channel-cut crystals acting as polarizer and analyser, respectively. Not shown is the telescope in between both and the optical laser responsible for polarising the vacuum. Right: Extinction curve around $0.01^\circ$ of the crossed-polarizer position. At a few data points, the corresponding detector signal is displayed: In crossed-polarizer position not a single photon reaches the detector. To the level tested in this experiment, the polarizers are perfect.}
    \label{fig:conv-exp}
\end{figure}

The first experiment at a free-electron x-ray laser was performed at the High Energy Density (HED) instrument of the European XFEL. The silicon 400 reflection and 6 reflections in the CC were used. A polarisation purity of $8\times 10^{-11}$ was achieved; see Fig.~\ref{fig:conv-exp} for an illustration. It should be emphasised that this number is {\it not} determined by the quality of the polarizers but by the photon flux of the XFEL. In other words: $8\times 10^{-11}$ is the {\it upper limit} for the actual extinction ratio; the extinction curve is in fact compatible with {\it perfect} polarisation \cite{Schulze:2022}. It should also be emphasised that the limited photon flux was due to limited beam time, the XFEL running in SASE mode, and other factors, i.e. they do not constitute principal road blocks. This is highlighted by the fact that the present record polarization purity of {$(1.4\pm 0.9)\times 10^{-11}$} was achieved at a 3-rd generation synchrotron facility \cite{Marx:NJP22}. The theoretical limit for the polarisation purity as determined by the finite laser beam divergence is on the order of $10^{-14}$! It has yet to be shown, whether this limit can be reached or whether other effects kick in. 

Another bottle neck are the lenses of the x-ray telescope. Due to a refractive index slightly smaller than unity in the x-ray regime, stacks of concave lenses must be used, so-called compound refractive lenses (CRL). Irrespective of that, it is known that x-ray diffraction in crystals can lead to birefringence. It is also known that metals, including Beryllium, the preferred material for x-ray lenses, have a micro-crystalline structure. Accordingly, the question has been whether this is indeed a limiting factor and, if so, how to bypass it. The Jena group has addressed both. In short, the answer is that there is little if any hope that Beryllium, which is attractive because of high transmission, can be used. Already $500\,\upmu{\rm m}$ of Beryllium degrade the polarisation purity to the level of $10^{-6}$ \cite{Grabiger:APL20}!

One alternative is the use of amorphous lens materials. Two options were tested so far: CRLs made from glassy carbon and lithographically produced CRLs made from the photoresist SU-8. The disadvantage of the latter is that the production method only allows one-dimensional focusing. Therefore, two SU-8 CRLs, rotated by $90^\circ$ with respect to each other, have to be used for focusing and recollimation. This reduces the transmission of the total setup. On the other hand, these lenses are of very good optical quality and allow close to diffraction-limited performance \cite{Grabiger:APL20}. For glassy carbon, the situation was exactly the opposite. Meanwhile, better components have become available and tests will be carried out soon. 

Still another alternative, on first glance maybe paradoxically, are lenses made from perfect crystals, specifically diamond. They can be rotated around the optical axis such that they exhibit no birefringence. Also this option will be tested soon.

The conclusion is that the conventional scenario is still very promising and could indeed be the fastest way to prove vacuum birefringence. Essentially, it must be shown that the polarimeters can  achieve an extinction ratio of the order of $10^{-12}$. So far, there is no evidence whatsoever that this might not be the case.

\subsection{Dark-field scenario}\label{subsec:DFexperiment}

\begin{figure}[h!!]
    \centering
    \includegraphics[width=1\linewidth]{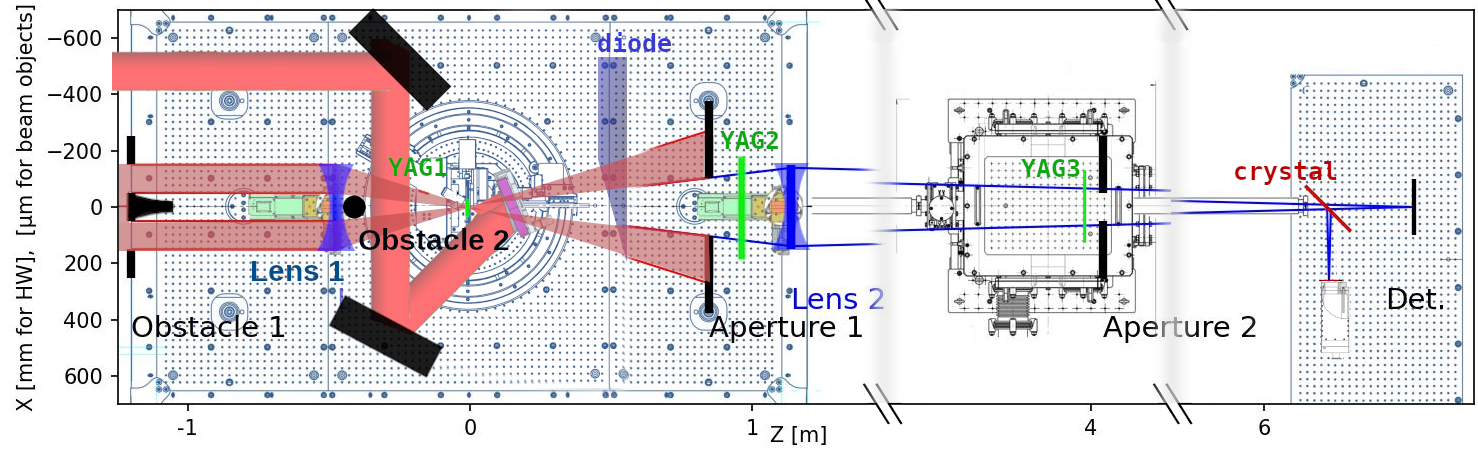}
    \caption{Experimental setup for the dark-field proof-of-principle experiment at the HED instrument of the European XFEL. For completeness, here also the RelaX beam path for the counter-propagating geometry is indicated.}
    \label{fig:Darkfield-exp}
\end{figure}

The setup described in Sec.~\ref{subsec:darkfield} and schematically shown in \figref{fig:DarkField} was adapted to fit the experimental conditions of the HED instrument of the European XFEL for the proof-of-principle experiment, as can be seen in \figref{fig:Darkfield-exp}. The experiment is designed to be used with XFEL photon energy of $8766\,{\rm eV}$, so that a 440 plane of a Germanium crystal can be used as a polarisation analyser. The XFEL beam is coming from the left side of the picture, entering the IC1 experimental chamber about $-1.2\,{\rm m}$~upstream from the interaction point. The whole setup is in vacuum chambers up till the kapton window located before detectors, at about $5\,{\rm m}$ downstream from TCC. At the entrance point ($-1.1\,{\rm m}$), obstacle O1 is located, which is made of a vertical wire-like structure with a diameter of $125-150\,\upmu{\rm m}$ to block the central part of the beam. The focusing lens (L1) is located $47\,\trm{cm}$ upstream from focus, and consists of 12 Beryllium lenses with a $50\,\upmu{\rm m}$ central radius of curvature. The Beryllium lenses have a $400\,\upmu{\rm m}$ diameter, which is the limiting factor for the incoming beam size. Just after the lens stack, obstacle O2 is located, made of simple wire with a size matching O1. A pinhole of about $50\,\upmu{\rm m}$ diameter is placed exactly in the focus, which prevents radiation scattered on the first lens from propagating further. The image of O1 projected by the lens L1 is at the position $85\,{\rm cm}$ downstream from the focus, where the aperture A1 is located. The scattering on the edges of these slits is a critical factor for the success of the setup, therefore several possibilities for how to limit the beam are being pursued, as described further. The second lens stack (L2), consisting of 6 lenses of the same type as lens 1, is located $114\,{\rm cm}$ downstream from focus, to image the focal plane onto the detector plane. An intermediate aperture (A2) is located in a separate chamber at the image of obstacle O2, i.e. about $4\,{\rm m}$ downstream of focus. Six to seven metres downstream, at the end of the experimental hutch, is where the detector bench is located. This can either host detectors in the direct beam, which is useful for initial alignment and test or can be used for measurement of the polarisation distinction. Alternatively, the analyser setup hosting 2 detectors and Ge crystal can be put in the hutch, as described below.

In order to align and monitor the x-ray beam, the diode screen with the photodiode, as well as several fluorescent screens are located along the beam propagation axis. For the ultimate discovery experiment the ReLaX optical beam will counter-propagate with the XFEL beam, focussed down by an $f/1$ off-axis parabolic mirror (OAP).In this experiment the ReLaX beam is expected to deliver $300\,{\rm TW}$  in $25\,{\rm fs}$, and be focussed down to a focal spot FWHM diameter $1.3\,\upmu{\rm m}$. This focusing scheme, beam routing, and corresponding diagnostics are yet to be developed.

There are two key parameters to evaluate the quality of the Dark-Field setup.
The shadow factor $\cal S$ was already defined in the theory, \secref{subsec:darkfield}. The definition from the experimental point of view is that it is the ratio of x-ray photons in the active area of the final detector to the number of x-ray photons entering the chamber if there is no scattering at the focal spot involved.

The aim of the design of a good setup is then to minimise the $\cal S$ factor while keeping the overall transmission sufficient. Therefore a transmission factor $T$ is defined to indicate a chance of producing \emph{and} detecting vacuum birefringence scattering processes. As the angular distribution of vacuum birefringence scattered photons is not known at this point, we define $T$ simply as $T=T_0 \times T_1$, where $T_0$ is the transmission from entrance to the chamber towards the focal position (therefore indicating how many photons will be interacting with the field), and $T_1$ is a transmission from the focal point towards the active area of the detector when both obstacles O1 and O2 would be removed - therefore indicating chance that a scattered photon will be detected. By the \emph{active area} of the final detector, we mean an area containing the image of the focal spot, therefore an area where the vacuum birefringence scattering can be detected. Distinguishing this area, which might be smaller than $10\,\upmu{\rm m}$ from the rest area of the detector is an important part of the setup.

Therefore it can be said, that the factor $T$ represents the signal in the experiment, while the shadow factor $\cal S$ represents the noise. The standard way of optimisation would be to maximise the signal-over-noise ratio, represented by the fraction $T/{\cal S}$. However, as will be shown later, this would lead to prohibitively low signal values (effectively closing up the apertures), providing zero measured photons. Therefore a measure of $T^6/{\cal S}$ was identified as an optimal quantity to be maximised, or less exactly, minimising the $\cal S$ factor while keeping the $T$ at a reasonable $10\%$ level.

\subsubsection{Diffractive simulations} \label{sec:diffractive-simulations}
\begin{figure}
    \centering
    \includegraphics[width=1\linewidth]{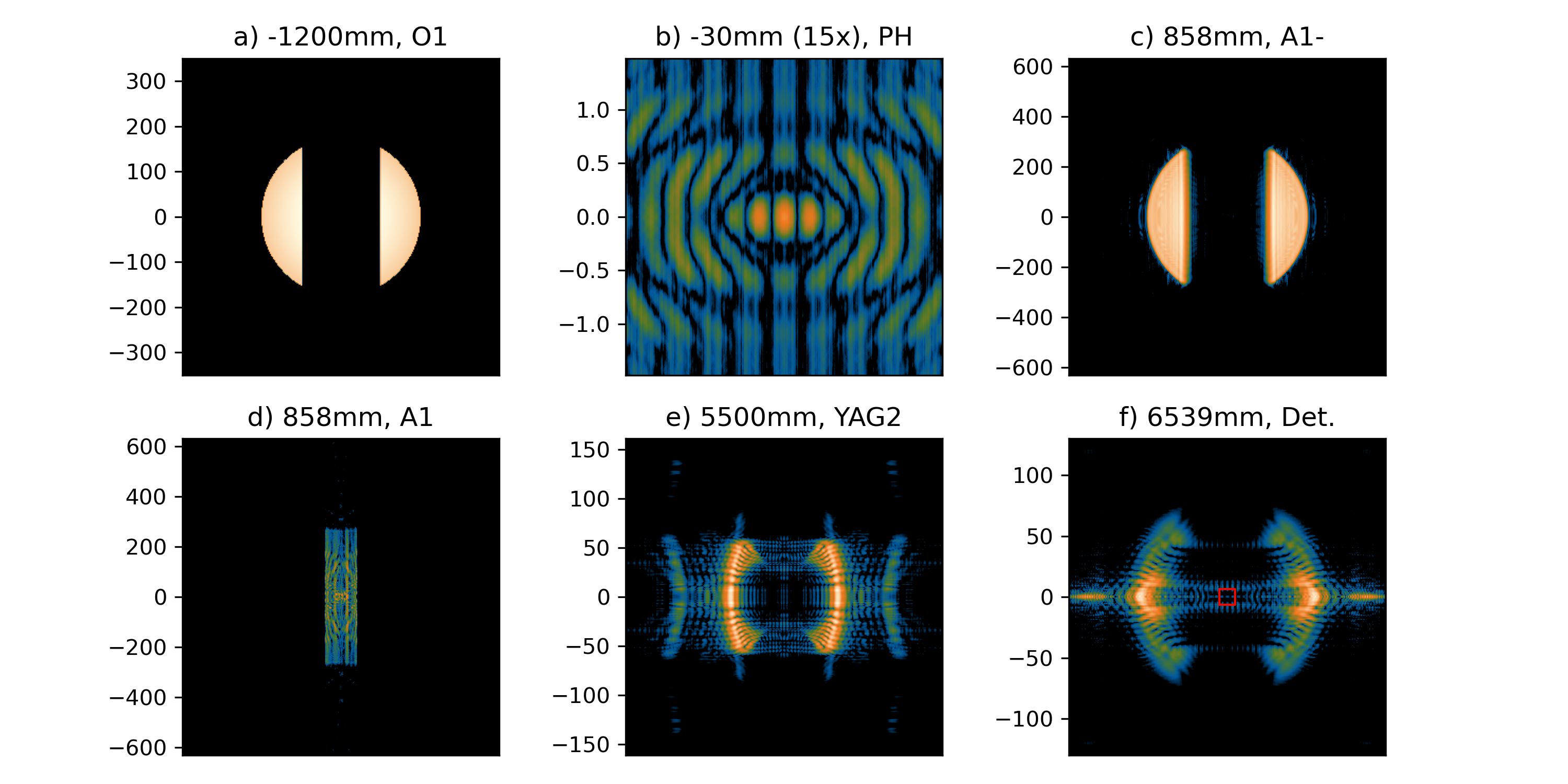}
    \caption{Result of the diffractive simulation. The sub-figures show the 2D intensity profiles of the beam along the beam path at various positions: (a) just behind the first obstacle, (b) at the pinhole position, close to beam focus, (c) before aperture A1, (d) behind aperture A1, (d) at an intermediate position, and (f) at the detector, with a red square indicating the area into which the signal scattered at focus would be imaged. The axes are in units of $\upmu{\rm m}$ and the colour scale is logarithmic over 3 orders of magnitude.}
    \label{fig:diffractive}
\end{figure}

In order to design and optimise the x-ray beam path and all its elements, the whole setup was simulated by a complex simulation using the LightPipes framework \cite{lightpipes}, which is designed to propagate a coherent beam where diffraction is essential. Most of the simulations presented here have a $700\,\upmu{\rm m}$ box size, resolution of $47\,{\rm nm}$ and therefore $15000 \times 15000$ simulated points. This was tested to be sufficient to avoid numerical imprecision for given cases. All obstacles and apertures are modelled as 2D maps of thickness, i.e. their exact 3D shape is neglected. The thickness is converted into transmission and phase shifts by using the online Henke tables  \cite{henke}. The full result of a typical simulation is shown and described in \figref{fig:diffractive}

\begin{figure}
    \centering
\includegraphics[width=0.5\linewidth]{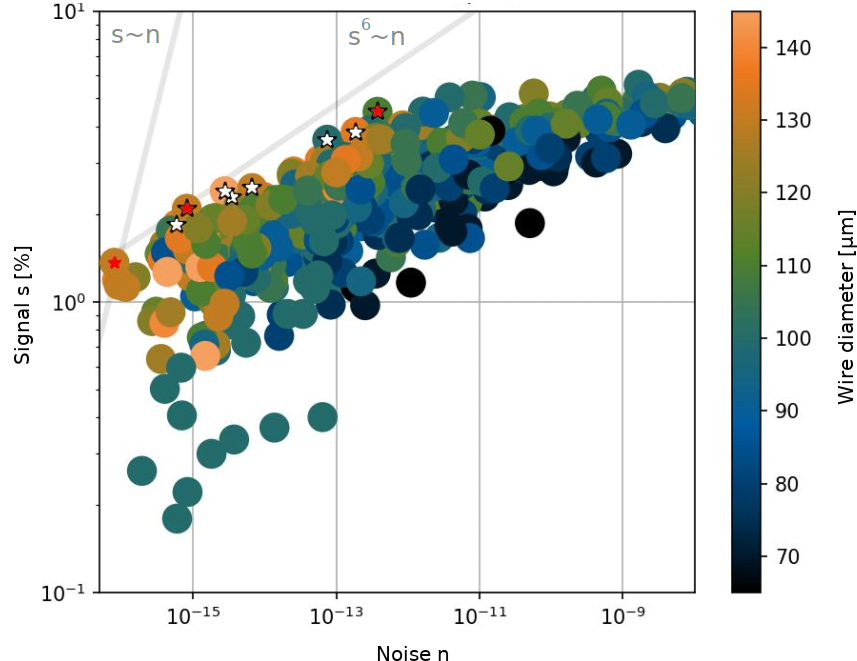}
\vspace*{1mm}
\caption{Example of optimisation of the experimental parameters. The horizontal axis is the shadow factor, the vertical one is the signal transmission factor. The diameter of the wire (obstacle O1) is encoded in the colour of the points.}
\label{fig:wireopt}
\end{figure}

An example of parameter optimisation is shown in \figref{fig:wireopt}. Here each point represents one set of parameters defining the experimental geometry, and its colour corresponds to the diameter of the obstacle O2 in $\upmu{\rm m}$. 
It can be seen that optimal obstacles are rather large with a diameter of around $140\,\upmu{\rm m}$. It is very useful to see that the typical optimal simulations are following the trend $n \approx s^6$, where $n$ corresponds to the noise of the measurement (which is proportional to the shadow factor), and signal $s$ is representing the possibility to detect a VB signal photon. That means that a simple optimisation of the signal-to-noise ratio would lead to prohibitively small signals. Therefore in further considerations, an approach to minimise the noise while keeping the signal on the $10\%$ level was used.

\subsubsection{Obstacles}
As basic types of obstacles will be round wires. As the diffraction and scattering on the edges of the wires are critical, their surfaces shall be polished to high surface quality, either by FIB or electrochemical etching. Commercially available diameters of $200\,\upmu{\rm m}$ and $175\,\upmu{\rm m}$ will be used and simulations have confirmed that the diameter of O2 shall be smaller than that of O1.

Alternatively, a custom-made microfabricated shape was developed to minimise the diffraction.  This so-called \emph{trumpet} shape exhibits a quadratic increase of thickness as a function of decreasing distance from the axis, which transfers to an exponential increase of its opacity. Therefore, as the tip of such a shape is very thin, given by the manufacturing possibilities in the order of a single $\upmu{\rm m}$, the diffraction on such a tip is very limited. On the other hand, the close-to-transparent edge might cause a significant refraction effect, which diverts the photons away from the axis, therefore such photons do not propagate further downstream and do not contribute to the noise on the detector.

\begin{figure}
    \centering
    \includegraphics[width=0.8\linewidth]{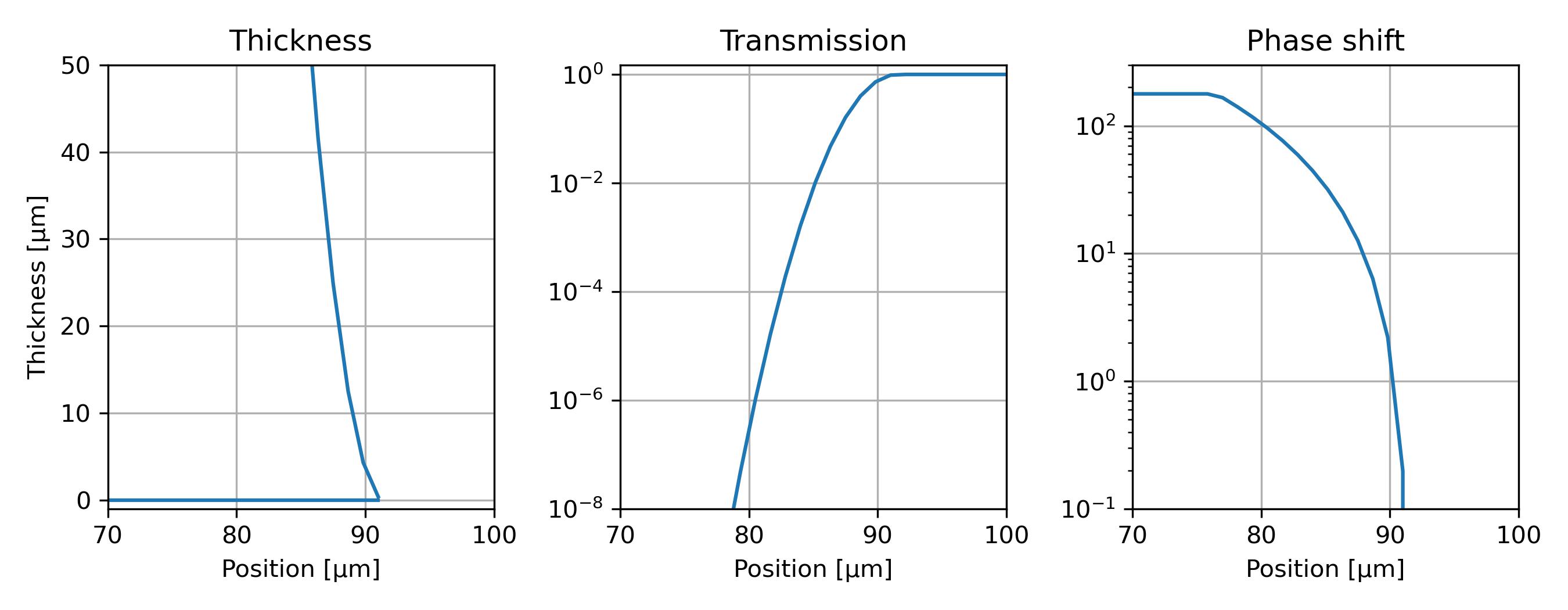}
    \caption{Design of a microfabricated obstacles. The left pane shows their shape (thickness as a function of position perpendicular to beam), the other two panes show the transmission and phase shift induced to the XFEL beam, the combined effect of which is to deflect the beam on the edges rather then to scatter it.}
    \label{fig:trumpets}
\end{figure}

A variant of such object is depicted in \figref{fig:trumpets}. The left pane shows the thickness profile as a function of radial distance. The profile is symmetrical, only positive values are shown and are in a wire-like geometry, i.e. extended over the not shown dimension.
The second and third subplots show the calculated absorption and phase shift caused by the obstacle.

\subsubsection{Performance evaluation}
\begin{figure}
    \centering
    \includegraphics[width=\linewidth]{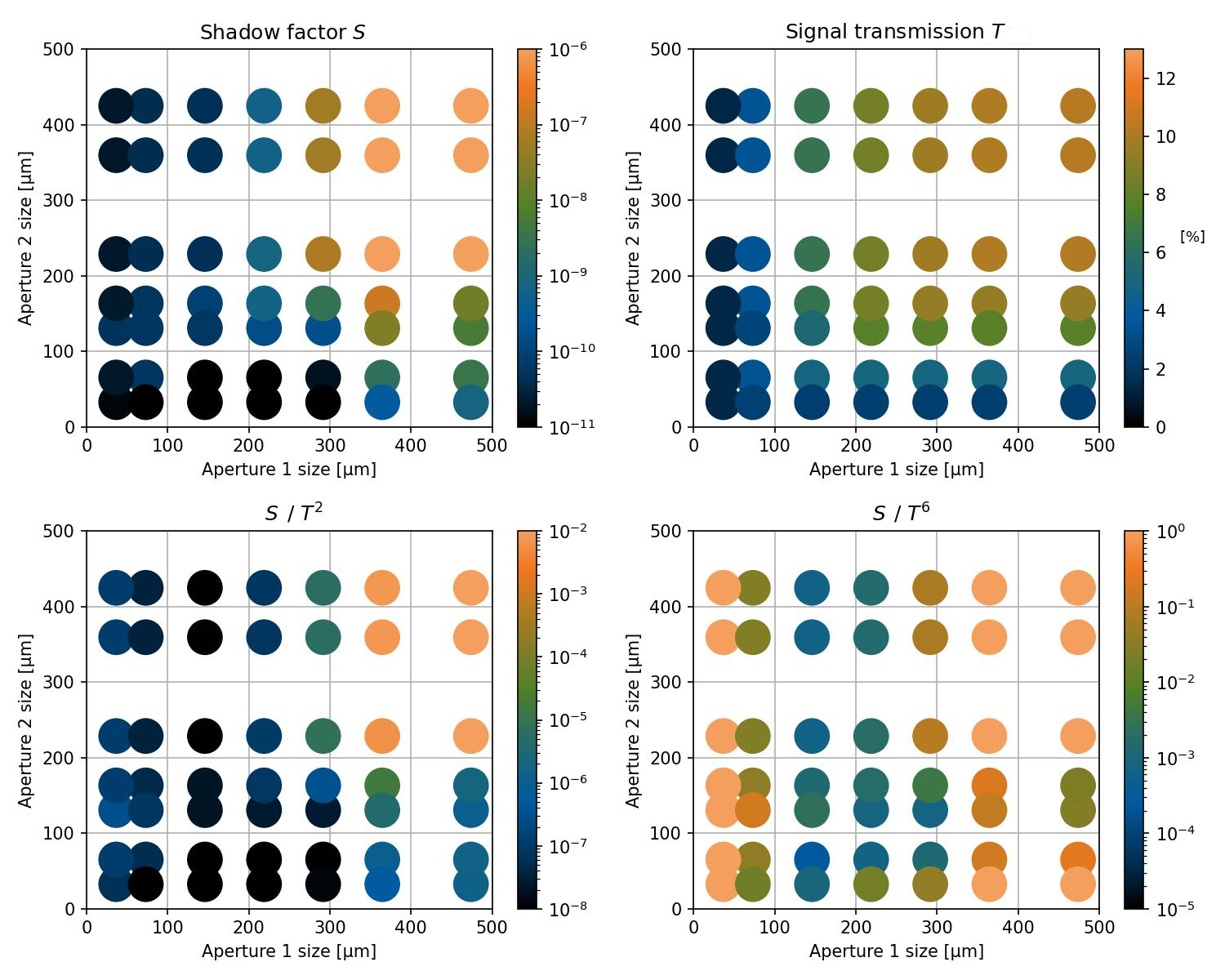}
    \caption{Diffraction simulation of various openings of apertures A1 and A2 slits while using the wires as obstacles. The first two figures show the simulated shadow and transmission factors, while the bottom figures show derived ${\cal S}/T^2$ and ${\cal S}/T^6$ factors, which are considered for optimisation. Minimum values of the latter factors are desirable for our purpose.
    }
    \label{fig:A1A2}
\end{figure}

Once a set of all beamline elements is given, it is still not trivial to find the optimal settings for the opening of both apertures A1 and A2. Opening of each aperture has its optimal value, which could not be independent of the other one. In general, if an aperture is open too wide, a significant amount of scattered radiation can go through, while if it is closed too much, the signal transmission is decreased and eventually, additional scattering on the slit edge can be produced. Therefore a set of simulations for various openings is performed for a given set of components. Figure \ref{fig:A1A2} shows the simulated $\cal S$ and $T$ factor in the top row, and the derived quantities ${\cal S}/T^2$ and ${\cal S}/T^6$ in the second row, for the case that both O1 and O2 are tungsten wires with respective diameters of $200\,\upmu{\rm m}$ and $175\,\upmu{\rm m}$. The best performance, i.e. the minimum value, of ${\cal S}/T^6 = 2\times10^{-4}$ is achieved for openings of $150\,\upmu{\rm m}$ and $80\,\upmu{\rm m}$, respectively. Different size of A1 increases this factor significantly, while A2 does not have that high influence, showing that the majority of signal reduction is done on the O1 -- A1 pair, while the 02 -- A2 pair plays rather a minor role.

\begin{figure}
    \centering
    \includegraphics[width=\linewidth]{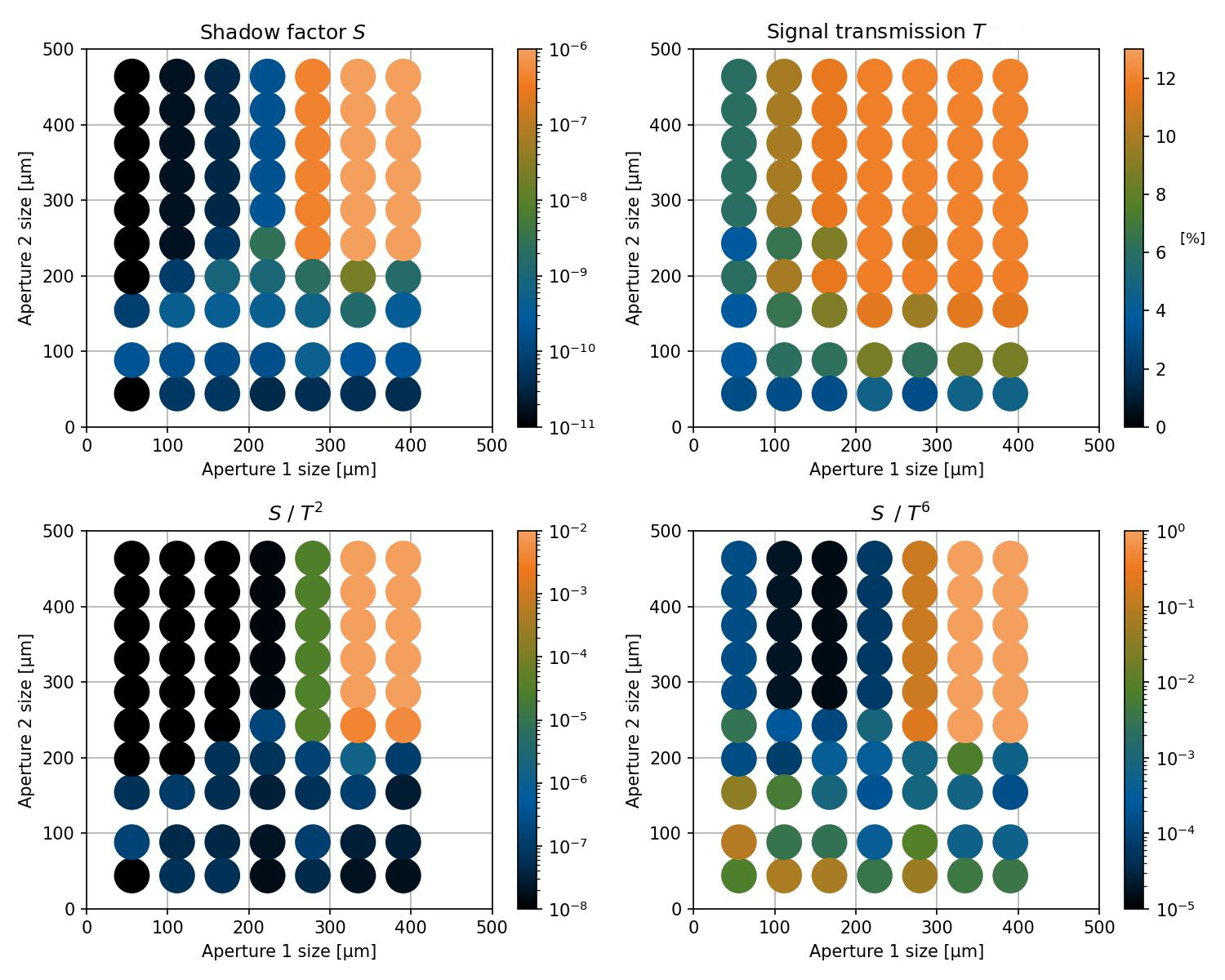}
    \caption{Diffraction simulation of various openings of the apertures A1 and A2 slits while using the \emph{trumpet} as obstacle O1 and the phase-corrected aperture as A1.}
    \label{fig:A1A2trumpet}
\end{figure}

This effect is seen even stronger in simulations with optimised O1 -- A1 components, as seen in \figref{fig:A1A2trumpet}. Here, the trumpet shape for O1 is used, and A1 is used with the soft edges of W slits with a plastic phase corrector. By this approach,  we improve the best-simulated value for $\cal S$ by a factor of $\approx30$, and the value for ${\cal S}/T^6$ by a factor of $\approx20$. What is interesting is to see that this value is for A2 size of $300\,\upmu{\rm m}$, and is essentially not changing if the aperture is opened more, showing that the cleaning on the first set of components is so superior, that the O2 -- A2 set does not produce any improvement.

\subsubsection{Ge analyser}\label{subsubsec:Ge}
As mentioned above, the advantage of the Dark-field scheme relies on combining two ways to distinguish the LbL scattered photons from the non-scattered (direct) XFEL photons, the angular scattering and the change of polarisation. The set of obstacles and apertures is performing the angular selection, while the second one is yet to be done by a polarisation analyser.  The main requirements when designing an analyser for these setups were identified as:
\begin{enumerate}    
    \item Bandwidth covering the majority of the XFEL spectrum.
    \item Possibility to measure both polarisation states.
\end{enumerate}

As we are relying on crystal-based polarisers, where the setting of Bragg angle to $45^\circ$ ensures reflection of only the $s$-polarisation, we have to admit that to our knowledge, all considered crystal reflection in their default configuration have insufficient bandwidth, i.e. below or at $0.2\,{\rm eV}$, while the bandwidth of self-seeded XFEL is in the order of $1\,{\rm eV}$.

\begin{figure}
        \centering
        \includegraphics[width=0.8\linewidth]{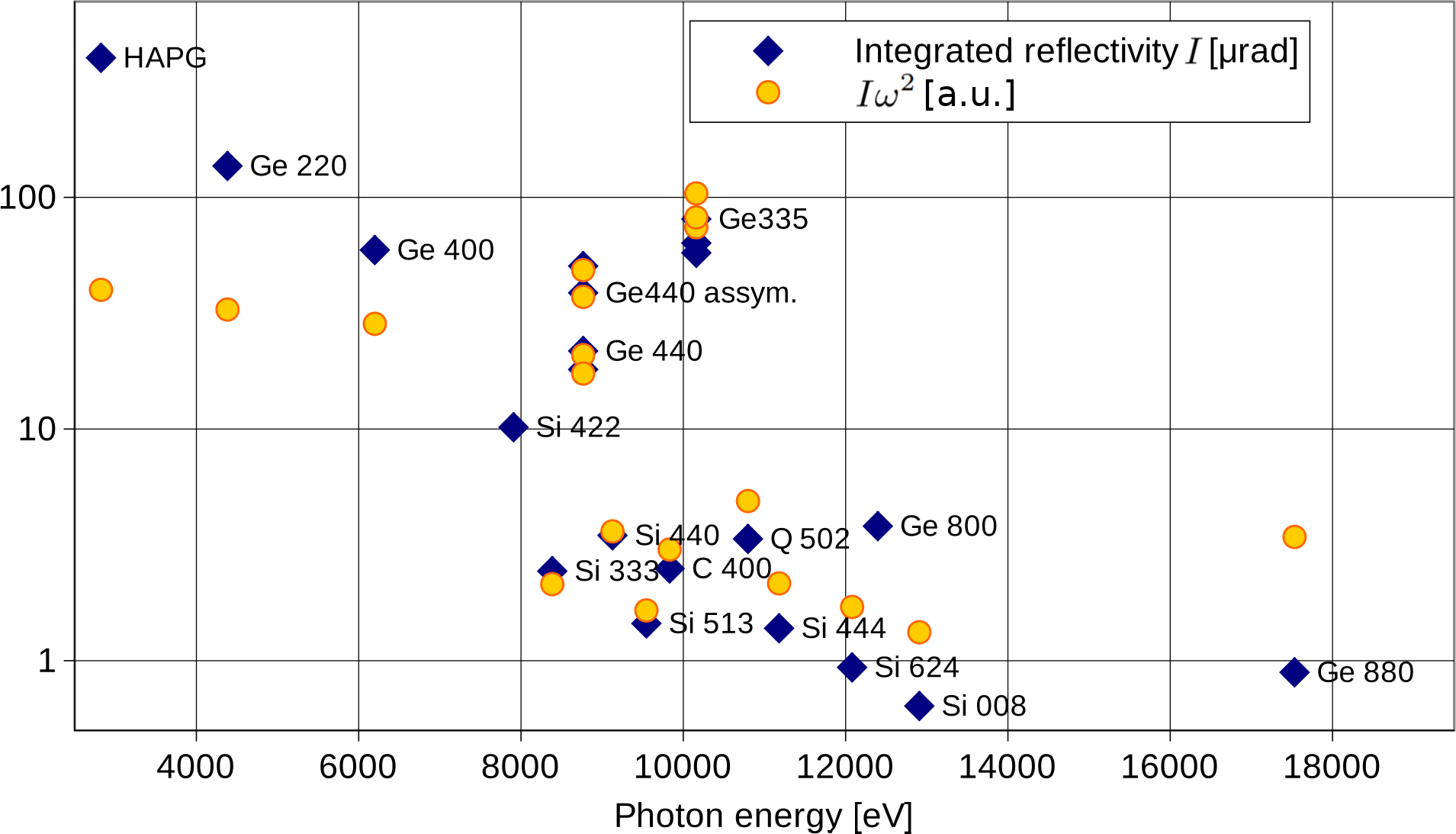}
        \vspace{1mm}
        \caption{Integrated reflectivity of considered crystal cuts.}
        \label{fig:crystals}
    \end{figure}

This is pushing the design to use asymmetrical cut crystals, where the angle of incidence is not equal to the Bragg angle. In this configuration, the bandwidth can be significantly improved without losing much of the peak reflectivity, therefore also the integrated reflectivity is increased. A summary of the performance of considered or common crystals is shown in \figref{fig:crystals}, where the integrated reflectivity is plotted as a function of photon energy, where the given plane can serve as a polariser. However, it is important to realise that for the success of the vacuum birefringence experiment, the quantity $I\omega^2$, where $I$ is the integrated reflectivity (approximated as peak reflectivity times Darwin width) and $\omega$ is the probe photon energy, is a good indicator for a suitable crystal, as the number of scattered photons scales as $\omega^2$. The crystal cuts ending with \_A or similar are those used with asymmetrical cuts with various degrees of asymmetry. That clearly shows that either the Ge 440 or Ge 335 in asymmetrical cuts are the crystals of choice. From those the Ge 440 was chosen, which fixes the experimental constraint to $8766\,{\rm eV}$. Various cuts and their integrated reflectivities are summarised in \tabref{tab:germanium}. Theoretically, it is clearly better to employ shallower incidence, which however requires a larger crystal surface with perfect quality. Therefore various crystals with different cuts are being used for the upcoming experimental campaign, to find which will perform best in given experimental conditions.

\begin{table}
    \centering
    \begin{tabular}{|c||c|c|c|c|}
    \hline
         Case & Angle between surface and 440 plane [$^\circ$] & Darwin width [$\upmu{\rm rad}$] & Bandwidth [$\rm eV$] & Integrated reflectivity [$\upmu{\rm rad}$]  \\
         \hline\hline
         standard (110) & $0$ &  $21$ & $0.18$ & $18$\\
         \hline
         111 & $35.25$ & $45$ & $0.39$ & $39$ \\
         \hline
         001 +$5.0^\circ$ miscut & $40$ & $62$ & $0.54$ & $52$\\
         \hline
         001 +$2.5^\circ$ miscut & $42.5$ & $88$ & $0.77$ & $69$\\ 
         \hline
    \end{tabular}
    \caption{Parameters of different cuts of Ge 440 crystal reflection.}
    \label{tab:germanium}
\end{table}

The second condition - possibility to measure both polarisation states - could be easily fulfilled by employing a thin, transmissive, crystal, as is done in other spectrometers already used e.g. at the HED instrument \cite{ZastrauHED}. However, two issues arise if that would be used in the asymmetric  (shallow Bragg angle) geometry: the crystal would have to be extremely thin, and its surface would have to be very precise along a large area. A combination of these two constraints leads us to the conclusion that such precision is not feasible. Therefore, the so-called Baronova configuration \cite{Baronova2008} is employed, which utilises a thick crystal in reflective geometry, but in such a geometry that both polarisation states are reflected on different planes, therefore in different directions. This setup will be adopted at XFEL such that the native XFEL polarisation (horizontal) will be reflected above the beam, while the polarisation-flipped photons will be reflected horizontally to another camera. The reflectivities for both beams shall be in the predicted value of about $50\,\upmu{\rm rad}$.

\section{Conclusions and Outlook}
\label{sec:concl}

The HED-HIBEF facility at the European XFEL offers a unique opportunity for the 
exploration of fundamental properties of the ground state of nature. From a 
microscopic viewpoint, HED-HIBEF has access to a parameter range of 
sufficiently large centre of mass energy and photon number density to  
measure the light-by-light scattering cross section which is induced by 
quantum fluctuations of QED degrees of freedom. From an effective macroscopic 
viewpoint, HED-HIBEF can create sufficiently strong fields to discover a 
nonlinear response of the quantum vacuum, violating the classical superposition 
principle and providing the vacuum with medium-like properties.

The experiment planned by the present collaboration aims at the 
discovery of vacuum birefringence, taking advantage of several recent developments: in 
addition to the HED-HIBEF facility reflecting the progress in generation and 
control of ultra-intense near-IR and x-ray pulses, our concept draws from the 
evolution of x-ray optics, most prominently high-purity polarimetry, as well as 
novel theoretical tools for the prediction of quantum signatures generated by 
generic spatiotemporal pulse distributions. 

Because of the diminutiveness of the expected signal, our collaboration 
envisages several prospective scenarios designed to isolate the nonlinear 
quantum signature from the expected large linear background. 
First beam time will be used to pursue the dark-field scenario where the discovery 
potential hinges on the shadow factor $\mathcal{S}$ in combination with x-ray 
polarimetry as a measure for the quality of background suppression. 
Currently available facility parameters together with theory predictions, diffractive 
simulations and practical feasibility suggest this scenario for a first step. 
However, unforeseen obstacles or improvements in 
other crucial experimental parameters may subsequently give preference to 
one of the other scenarios brought forward by our collaboration.

The first discovery of vacuum birefringence would be a landmark for several
reasons: Within QED, vacuum birefringence and the Cotton-Mouton constant of 
the vacuum is a genuine prediction similar to phenomena such as the Lamb shift 
or the anomalous magnetic moment of the electron. In the same manner as the 
latter has been measured to increasing precision, vacuum birefringence should 
be studied in all detail, because precision experiments provide stringent 
tests of our understanding of nature, each one coming with its own discovery 
potential for unexpected deviations. 

Vacuum birefringence gives direct access to the HE effective action. For 
instance, the dark-field scenario and the three-beam set-up
can measure the two Heisenberg-Euler 
coefficients $c_1$ and $c_2$ separately. This makes these  types of experiments an 
ideal laboratory for investigating effective actions more generally. The fact that
effective actions are used for effective field theories in modern 
physics ubiquitously, e.g., in particle physics 
\cite{Weinberg:1978kz,Georgi:1993mps}, solid-state physics 
\cite{Zhang:1988wy,Shankar:1993pf}, cosmology 
\cite{Cheung:2007st,Carrasco:2012cv}, searches for hypothetical new physics  
\cite{Brivio:2017vri,Fitzpatrick:2012ix} and even in quantum gravity 
\cite{Buchbinder:1992rb,Donoghue:1994dn}, with the HE action establishing this 
important concept for the first time makes systematic studies so desirable. In 
the present context of strong-field QED, the effective action can be studied 
below and -- for increasing probe photon energy -- near the mass threshold.

As the nonlinear response of the vacuum is a result of quantum fluctuations of 
all interacting degrees of freedom, the Cotton-Mouton constant of the vacuum 
$k_{\text{CMV}}$ is not only determined by electron-positron fluctuations. To 
lowest order in the coupling constant, all charged degrees of freedom contribute 
as a matter of principle. However, since $k_{\text{CMV}}\sim 1/m^4$, the 
contributions of the next-to-lightest charged particles, muons, and charged 
pions, with masses above $100\,{\rm MeV}$ are suppressed by $9-10$ orders of magnitude, presumably interfering with the QED contributions beyond the 4-loop level. In contrast to high-energy collider-type experiments, the discovery potential of 
quantum vacuum experiments extend to the regime of small masses but possibly 
weakly coupled degrees of freedom \cite{Gies:2008wv,Jaeckel:2010ni}. 

In fact, already the first cavity-based experiment, BFRT, searching 
for vacuum birefringence as well as vacuum dichroism produced limits on 
hypothetical small-mass scalar or pseudoscalar degrees of freedom 
\cite{Semertzidis:1990qc} and their potential coupling to photons.
By now, the set of hypothetical particles -- often motivated as dark-matter candidates -- that can be searched for in quantum vacuum experiments includes  (pseudo-)scalar axion-like particles, scalar or fermionic minicharged particles,  or additional light vector bosons, see, e.g., \cite{Ahlers:2007rd,Ahlers:2007qf}  for a corresponding analysis of the published PVLAS data.
In principle, probe photons exposed to a strong field can also mix with neutrinos or gravitons 
inducing a tiny standard-model background for vacuum dichroism 
\cite{Raffelt:1987im,Gies:2000wc,Ahlers:2008jt}. 

The strongest bounds on the set of hypothetical particles typically come from 
astrophysics or cosmology, since such extra degrees of freedom can contribute 
to heat or photon transport and thus modify generic time or length scales 
subject to astrophysical observations. However, such bounds often rely on 
further (stellar or cosmological) model input 
\cite{Jaeckel:2006xm,baker13,Hoof:2021mld}, justifying independent purely 
laboratory-based experiments as provided by vacuum response measurements. 

The search potential of the vacuum birefringence measurement at HED-HIBEF 
depends on both the hypothetical model degrees of freedom as well as on the 
details of the measurement scenario. Compared to earlier experiments providing 
laboratory bounds such as PVLAS, ALPS, or OSQAR 
\cite{Ejlli:2020yhk,Ehret:2010mh,OSQAR:2013jqp}, several aspects are 
different: e.g., with reference to Eq.~\eqref{eq:deltasquared}, our experiment 
is designed to 
improve on the flip probability $\sim I_{\rm L}^2 
z^2\omega_{\rm X}^2$ by using an ultra-intense laser with large 
$I_{\rm L}$ and an XFEL with large $\omega_{\rm X}$, however, at the 
expense of a small interaction length $z$. Naively, the flip probability for 
axion-like-particles in certain regimes scales as $\sim I_{\rm L} 
z^2$, for minicharged particles as the QED result for birefringence at small frequencies and as 
$\sim I_{\rm L}^{2/3} z^2/\omega_{\rm X}^{2/3}$ for large frequencies where also dichroism becomes significant, and 
is even independent of $I_{\rm L}$ for vector bosons with kinetic mixing.

An advantage of HED-HIBEF using an XFEL probe beam is to provide access to 
mass scales up to the keV regime \cite{Villalba-Chavez:2016hxw,Evans:2023jpr}, 
where the bounds of \cite{Ejlli:2020yhk,Ehret:2010mh,NOMAD:2000usb} for 
axion-like particles are rather weak. (Note that bounds in the higher-mass 
regime  \cite{AxionLimits} stem from collider experiments  and often do not 
directly measure the axion-photon coupling but proceed via its embedding into 
the electroweak gauge group \cite{BaBar:2021ich}.) Also, enhancements due to 
resonance effects \cite{Dobrich:2012sw,Dobrich:2012jd,Villalba-Chavez:2016hxw,Evans:2023jpr} 
may increase the 
search potential considerably. We emphasise that further possible advantages 
arising from the use of, e.g., the dark-field scenario relying on the 
spatiotemporal structure of the field still remains to be explored.

Let us finally remark that experimental studies of nonlinear vacuum response to 
strong fields opens a new window on fundamental physics probing the parameter 
space of high amplitude rather than high energy. The scientific potential of 
this research area has yet to be realised. In a manner analogous to 
nonlinear media or (relativistic) plasmas for triggering optical 
phenomena used for a variety of applications, it is conceivable that the 
quantum vacuum will finally be put to use as the ultimate medium at highest 
intensities.

\bmhead{Acknowledgements}

We thank the ExtreMe Matter Institute EMMI at GSI, Darmstadt, for support via an EMMI Collaboration Meeting where this Letter of Interest has been initiated.
This work has been funded by the Deutsche Forschungsgemeinschaft (DFG, German Research Foundation) under Grants Nos.\ 392856280, 416611371, 416607684, 416702141, and 416708866 within the Research Unit FOR2783/2 and Project-ID 278162697 -- SFB 1242. AJM is supported by the project ``Advanced Research Using High Intensity Laser Produced Photons and Particles'' (ADONIS) CZ.02.1.01/0.0/0.0/16\_019/0000789 from the European Regional Development Fund (ERDF). We also thank Guido Zavattini (PVLAS) as well as Daniel Brandenburg and Frank Geurts (STAR) for their kind permission to reproduce Figs.~\ref{fig:PVLAS} and \ref{fig:STAR} (b).

\bibliography{bibliography}

\end{document}